\documentclass[11pt]{article}
\usepackage[T1]{fontenc}
\usepackage{lmodern}
\usepackage[margin=1in]{geometry}
\usepackage{amsmath,amssymb,amsthm,mathtools}
\usepackage{booktabs,array}
\usepackage{microtype}
\usepackage{tikz}
\usepackage{xurl}
\usepackage[colorlinks=true,linkcolor=blue!55!black,citecolor=blue!55!black,urlcolor=blue!55!black]{hyperref}
\hypersetup{pdftitle={Binary Belief Propagation Fixed Points: Finiteness, Rigidity, and Topological Complexity},pdfauthor={Yusuke Watanabe}}

\newtheorem{maintheorem}{Theorem}

\newtheorem{theorem}{Theorem}[section]
\newtheorem{lemma}[theorem]{Lemma}
\newtheorem{proposition}[theorem]{Proposition}
\newtheorem{corollary}[theorem]{Corollary}
\theoremstyle{definition}

\theoremstyle{remark}
\newtheorem{remark}[theorem]{Remark}
\numberwithin{equation}{section}
\DeclareMathOperator{\atanh}{arctanh}
\DeclareMathOperator{\asinh}{arsinh}
\DeclareMathOperator{\diag}{diag}
\DeclareMathOperator{\adj}{adj}
\DeclareMathOperator{\ord}{ord}
\newcommand{\R}{\mathbb R}
\newcommand{\C}{\mathbb C}

\newcommand{\ones}{\mathbf 1}
\newcommand{\abs}[1]{\lvert#1\rvert}
\newcommand{\leancommit}{4d0ae2b2}
\newcommand{\leanrepourl}{https://github.com/ywatanabex/bp-response-geometry-lean}

\title{Binary Belief Propagation Fixed Points:\\ Finiteness, Rigidity, and Topological Complexity}
\author{Yusuke Watanabe}
\date{9 November 2026}

\begin{document}
\maketitle
\begin{abstract}
A fixed finite binary Ising model has only finitely many admissible complex
belief propagation fixed points, even at degenerate parameter values.
Families of positive fixed points obtained by varying external fields
can nevertheless have nontrivial topology. We study this distinction through
the external-field projection. For a fixed field, we prove finiteness by
bounding the pole orders of the magnetizations on a hypothetical complex
solution curve. The generic degree on the support graph then bounds the
total local multiplicity. A graph has a single smooth real line as its full
positive response for every choice of positive couplings, base field, and
strictly positive sweep direction exactly when it contains no two
vertex-disjoint connected subgraphs of cycle rank at least two.
Two such subgraphs suffice to create a circular
component. For a plane obtained by varying two vertex fields, connected
simple bipartite cubic graphs with one uniform coupling and a universal
bound on the fixed fields realize every genus. The minimum number of
vertices grows logarithmically in the genus, by a matching algebraic
upper bound on Betti numbers. Finally, we certify the complete field
discriminant of an eighteen-vertex product model and recover its torus
component's genus from the counts of distinct roots on all field cells.
The three main theorems, including the logarithmic size bounds, are
formally verified in Lean~4 with Mathlib.
\end{abstract}

\section{Introduction and main results}
\label{sec:intro}

The external field provides a common parameter for several questions about
binary belief propagation (BP). At one field, we can ask how many fixed
points there are, including the complex ones allowed by the rational update.
Along a line of fields, we can ask whether all positive fixed points lie on
one continuation curve. With two independently variable fields, we can ask
which closed surfaces occur as components of the positive response. These
questions concern different inverse images of the same parameter
projection. On the positive real locus and at fixed coupling \(J\), this
projection is the external-field map
\(H_J\colon(-1,1)^V\to\R^V\) in the vertex magnetizations. The
complex fixed-field question is formulated in admissible message
coordinates.

The answers separate a fixed model from a family of models. A fixed finite
model has finitely many fixed points at every parameter, degenerate or not.
The solution set of the family obtained by moving the fields can
nevertheless have any topology, and the graph decides how much. Along a
positive-direction sweep the response is a single line exactly when the
graph has no two vertex-disjoint subgraphs of cycle rank two. Over a
two-field plane every genus occurs, already for connected simple cubic
graphs with one coupling value and bounded fixed fields, and the number
of vertices this requires grows only logarithmically in the genus.

\begin{figure}[tbp]
  \centering
\begin{tikzpicture}[x=.62cm,y=.50cm,font=\footnotesize]
  \begin{scope}
    \node[font=\small\bfseries] at (3.5,8.0) {All positive roots};
    \foreach \nx [count=\i from 0] in {3,5,7,9,7,5,3} {
      \foreach \ny [count=\j from 0] in {3,5,7,9,7,5,3} {
        \pgfmathtruncatemacro{\roots}{\nx*\ny}
        \pgfmathtruncatemacro{\cellshade}{8+\roots/2}
        \fill[blue!\cellshade!white] (\i,\j) rectangle ++(1,1);
        \node at (\i+.5,\j+.5) {\roots};
      }
    }
    \draw[step=1,black!50,thin] (0,0) grid (7,7);
  \end{scope}
  \begin{scope}[xshift=5.85cm]
    \node[font=\small\bfseries] at (3.5,8.0) {Roots on the torus};
    \foreach \nx [count=\i from 0] in {0,0,2,4,2,0,0} {
      \foreach \ny [count=\j from 0] in {0,0,2,4,2,0,0} {
        \pgfmathtruncatemacro{\roots}{\nx*\ny}
        \pgfmathtruncatemacro{\cellshade}{4+3*\roots}
        \fill[teal!\cellshade!white] (\i,\j) rectangle ++(1,1);
        \node at (\i+.5,\j+.5) {\roots};
      }
    }
    \draw[step=1,black!28,thin] (0,0) grid (7,7);
    \foreach \i in {2,3,4,5} {
      \draw[teal!65!black,semithick] (\i,2)--(\i,5) (2,\i)--(5,\i);
    }
  \end{scope}
  \foreach \shift in {0cm,5.85cm} {
    \begin{scope}[xshift=\shift]
      \foreach \tick [count=\i from 1] in {-c,-b,-a,a,b,c} {
        \node[below] at (\i,0) {\(\tick\)};
        \node[left] at (0,\i) {\(\tick\)};
      }
      \node[below] at (3.5,-.95) {\(h_1\)};
      \node[left] at (-1.25,3.5) {\(h_2\)};
    \end{scope}
  }
\end{tikzpicture}
  \caption{Open-cell root counts for the eighteen-vertex product model
  of Section~\ref{sec:diagrams}, the disjoint union of two nine-vertex
  models with one control field each.
  The field axes are schematic. The same thresholds
  \(-c,-b,-a,a,b,c\) cut each axis. The left panel counts all positive
  fixed points. The right panel counts those on the torus. The full
  discriminant consists of the twelve interior grid lines extended
  across the plane. The torus discriminant consists of the dark
  segments on the right. The outer cells extend to infinity.}
  \label{fig:discriminant}
\end{figure}
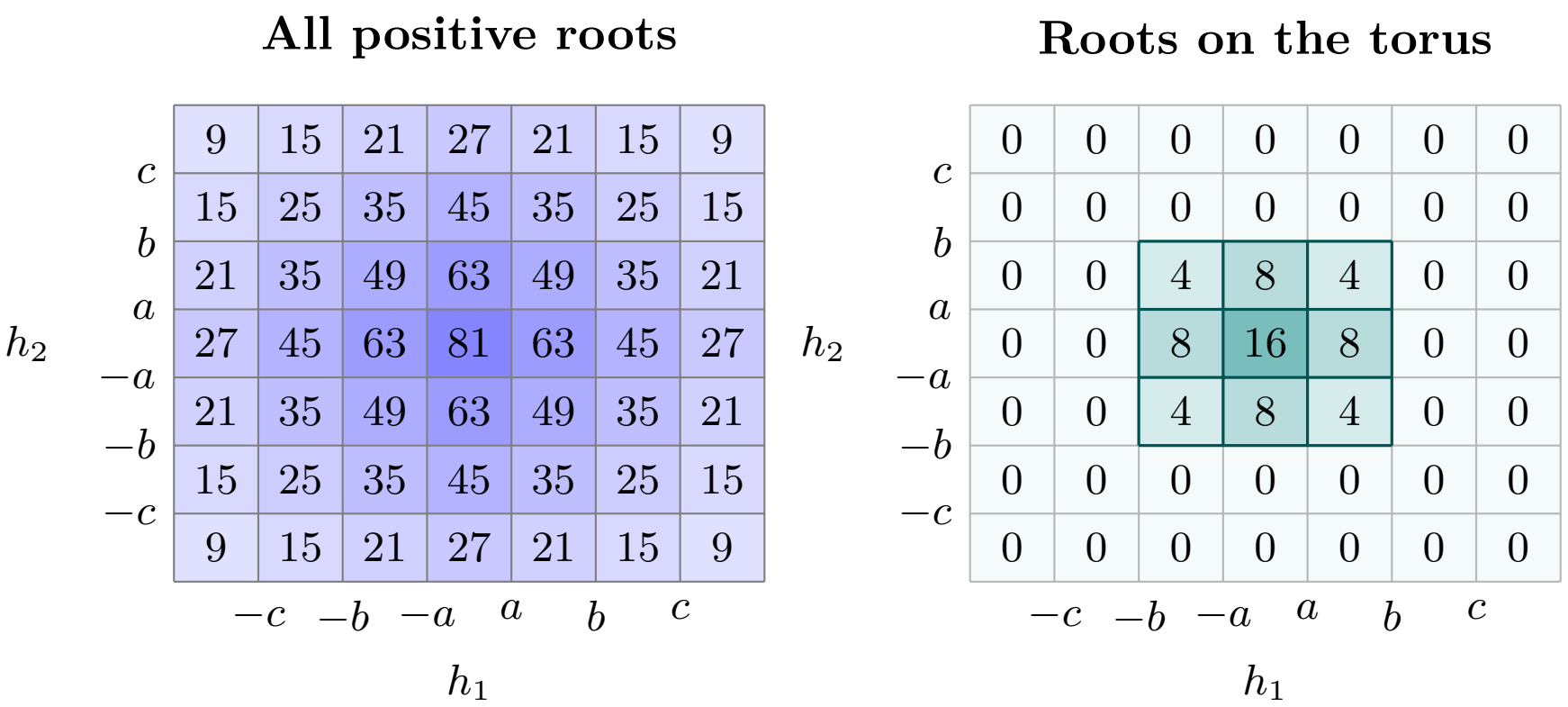

The distinction between a fiber and a response family is already visible
in Figure~\ref{fig:discriminant}. Each cell of the two-field plane has a
finite count of positive fixed points, while one component of the family
over that plane is a torus. The right panel counts only the points on that
component. Section~\ref{sec:diagrams} specifies the model and certifies
the counts and indices on the open cells, boundary lines, and intersections.
The model behind the figure is the disjoint union of two copies of a
nine-vertex model. Its product structure is what makes every count
certifiable by interval arithmetic. It is a certified illustration of
the phenomenon and not the construction of Theorem~\ref{thm:C}, which
produces connected graphs with one coupling value.

We work with finite loopless multigraphs unless simplicity is specified.
Parallel edges are distinct factors. The cycle rank is
\(\nu(G)=|E|-|V|+c(G)\), where \(c(G)\) counts connected components,
including isolated vertices. A complex fixed point is admissible
when its message ratios are finite and nonzero, all update denominators
are nonzero, and the vertex beliefs can be normalized. These conditions
also ensure that the edge beliefs can be normalized.

\begin{maintheorem}[Fixed-field finiteness]
\label{thm:A}
For every finite loopless multigraph and every fixed finite real choice
of couplings and external fields, the set of all admissible complex BP
fixed points is finite. This holds for signed and zero couplings and
includes degenerate roots.
\end{maintheorem}

The generic number \(N(G)\) of admissible complex roots is the graph
state sum \(Q(G)\) of \cite{counting}. That count concerns parameters in
a Zariski open set. It leaves open whether a special real parameter can
lie under a positive-dimensional component of admissible solutions, in
which case the fiber there would be infinite. Theorem~\ref{thm:A}
excludes this at every finite real parameter, including the degenerate
parameters at which roots collide. Its proof does not use the value of
\(Q(G)\). The generic count enters only afterwards, in
Corollary~\ref{cor:budget}: the sum of the local complex lengths at a
fixed real parameter is at most \(Q(G_0)\), where \(G_0\) is obtained by
deleting the zero couplings, and the positive roots consume at most
\(Q(G_0)-2^{\nu(G_0)}+1\) of that budget, where \(\nu\) is the cycle rank.

Theorem~\ref{thm:A} is the isolation hypothesis needed by methods that
locate BP fixed points as isolated roots of a polynomial system. Knoll,
Mehta, Chen, and Pernkopf \cite{homotopy} use polynomial homotopy
continuation, whose guarantees concern isolated complex solutions. For
finite real Ising models, Theorem~\ref{thm:A} shows that every admissible
root is isolated, including degenerate roots, so this hypothesis holds at
every parameter and not only generically. The theorem does not assert
that a particular numerical method finds all roots, nor that degenerate
roots are easy to compute. It settles the geometric hypothesis and leaves
the algorithmic guarantees to the method.

For a graph \(G\), let \(p_U(G)\) be the largest number of pairwise
vertex-disjoint connected subgraphs of cycle rank at least two.
The subgraphs may use any subset of the edges. For a nonempty graph,
a base field \(h^0\), and a direction \(v\in(0,\infty)^V\), define the
full positive response to be
\[
 \mathcal C_{J,h^0,v}
 =\{(m,s)\in(-1,1)^V\times\R:H_J(m)=h^0+sv\}.
\]

\begin{maintheorem}[Rigidity of positive-direction response]
\label{thm:B}
For a nonempty finite loopless multigraph \(G\), the following are equivalent.
\begin{enumerate}
  \item \(p_U(G)\leq 1\).
  \item For every finite choice \(J_e>0\), every \(h^0\in\R^V\), and every
    \(v\in(0,\infty)^V\), the full response \(\mathcal C_{J,h^0,v}\)
    is a connected real analytic one-manifold diffeomorphic to \(\R\).
  \item For every finite choice \(J_e>0\) and every \(h^0\in\R^V\), the response
    with \(v=\ones\) has no smooth circular connected component.
\end{enumerate}
For simple graphs these conditions are equivalent to excluding three
minors: \(B\sqcup B\), \(B\sqcup D\), and \(D\sqcup D\), with \(B\) the
bowtie of two triangles sharing a vertex and \(D=K_4-e\) the diamond.
\end{maintheorem}

Theorem~\ref{thm:B} should be compared with uniqueness results.
Watanabe's signed-graph classification \cite{signed} gives the graph
classes on which BP has a unique fixed point for every choice of coupling
magnitudes and fields. For positive couplings these are the graphs whose
connected components are trees or have one cycle, that is, the graphs
with \(p_U(G)=0\). Sufficient conditions for convergence of the iteration
to a unique fixed point, such as those of Mooij and Kappen \cite{mooij},
likewise describe parameter regions with one root. Theorem~\ref{thm:B}
asks a different question. It allows several positive roots at a field
and asks whether all of them stay on one curve as the field is swept in a
positive direction. The answer is a graph condition, and it is sharp: two
vertex-disjoint subgraphs of cycle rank two already allow a circular
component. The proof uses the order of roots in magnetization
coordinates and the signs of the tangent at turning points.

\begin{figure}[tbp]
  \centering
\begin{tikzpicture}[x=1cm,y=1cm,font=\footnotesize,>=stealth]
  \begin{scope}
    \draw[->] (-2.6,0) -- (2.8,0) node[below left] {\(s\)};
    \draw[->] (0,-1.9) -- (0,1.9) node[left] {\(m_i\)};
    \draw[thick,blue!60!black]
      plot[variable=\t,domain=-1.6:1.6,samples=80]
      ({\t*\t*\t-1.3*\t},{\t});
    \draw[dashed] (0.45,-1.85) -- (0.45,1.85);
    \foreach \t in {-0.8919,-0.3928,1.2846} {
      \fill[black] ({\t*\t*\t-1.3*\t},{\t}) circle (1.6pt);
    }
    \node[align=center] at (0,-2.45) {\(p_U(G)\le1\): one line,\\ three roots at the dashed field};
  \end{scope}
  \begin{scope}[xshift=7.2cm]
    \draw[->] (-2.6,0) -- (2.8,0) node[below left] {\(s\)};
    \draw[->] (0,-1.9) -- (0,1.9) node[left] {\(m_i\)};
    \draw[thick,blue!60!black]
      plot[variable=\t,domain=-1.5:1.5,samples=60]
      ({0.9*\t+0.35*\t*\t*\t},{\t});
    \draw[thick,teal!60!black] (1.9,0.6) ellipse (0.5 and 0.4);
    \draw[dashed] (1.9,-1.85) -- (1.9,1.85);
    \foreach \p in {(1.9,0.2),(1.9,1.0),(1.9,1.2857)} {
      \fill[black] \p circle (1.6pt);
    }
    \node[align=center] at (0,-2.45) {\(p_U(G)\ge2\): a line and\\ a circular component};
  \end{scope}
\end{tikzpicture}
  \caption{Schematic positive-direction responses, drawn as one
  magnetization coordinate against the sweep parameter \(s\).
  Left: for \(p_U(G)\leq1\) the full response is one analytic line.
  Its projection to \(s\) can fold, so a fixed field (dashed) can meet it
  in several positive roots. Right: for \(p_U(G)\geq2\) suitable
  parameters produce a circular component in addition to the line
  through the two strong-field ends.}
  \label{fig:one-line}
\end{figure}
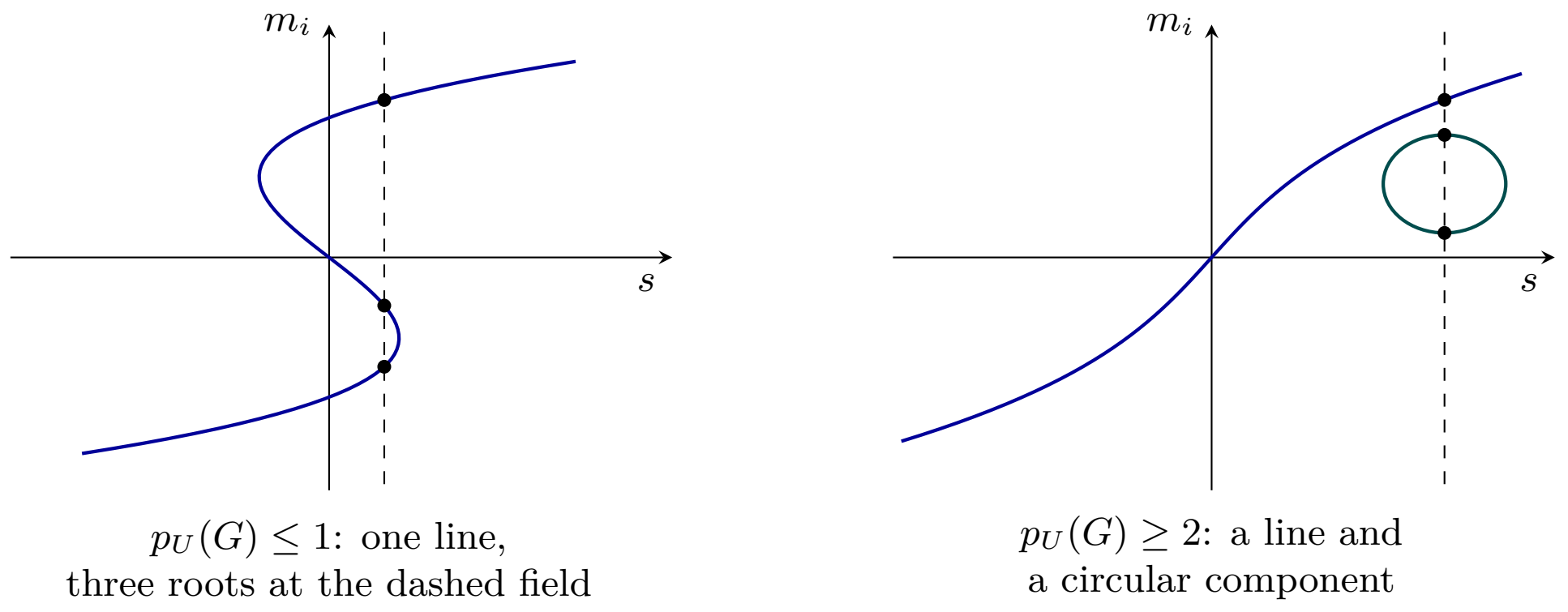

A single curve does not mean a unique root. The projection of the
response curve to the sweep parameter can fold, so one line meets a
fixed-field slice in several points (Figure~\ref{fig:one-line}). Within
the class \(p_U(G)\leq1\), the number of such points is unbounded as the
graph varies (Section~\ref{sec:rigidity}). For a fixed graph the number
of positive roots in a slice is bounded through Corollary~\ref{cor:budget}.

For two distinct controlled vertices \(D\subset V\), put \(W=V\setminus D\).
The response \(\mathcal S_D(J,h_W)=\{m:H_{J,W}(m)=h_W\}\) is taken with
both \(D\)-fields free. A compact connected component is regular if
\(D_mH_{J,W}\) has full row rank at every point on it. It is then a closed
orientable surface.

The third result says that, at one coupling value and under one bound
on the fixed fields, connected simple bipartite cubic graphs realize
every genus, and that the number of vertices needed grows only
logarithmically in the genus. The statement adds that a single graph
serves a whole range of genera, with explicit sizes.

\begin{maintheorem}[Genus and optimal size]
\label{thm:C}
Fix
\[
 J_0=\tfrac12\log(3+\sqrt{10}).
\]
There is a finite constant \(H_0\) such that for each integer \(k\geq 1\)
there is a connected simple bipartite cubic graph on \(366k+2550\)
vertices, with two designated controlled vertices and coupling \(J_0\)
on every edge, with the following property. For every
\(0\leq g\leq 3\cdot 2^k-1\), one can choose the noncontrolled fields
with \(\abs{h_i}\leq H_0\) so that the response to the two remaining
fields has a regular compact connected component of genus \(g\).
The graph, couplings, and control vertices for a given \(k\) are common
to all these choices. The constant \(H_0\) is independent of \(k\) and
\(g\). Once the fields on \(W\) have been chosen for a genus, they
are held fixed while the two fields on \(D\) vary independently.

With \(J_0,H_0\) fixed, the minimum number \(n_{\min}(g)\) of vertices
among all connected simple bipartite cubic graphs with uniform coupling
\(J_0\), two controlled vertices, and \(\abs{h_i}\leq H_0\) on \(W\)
that realize such a component of genus \(g\) is
\(\Theta(\log(g+1))\) for \(g\geq 1\). Genus zero is realized at
constant size.
\end{maintheorem}

That every closed orientable surface is a regular level set of a
polynomial map is elementary. The content of Theorem~\ref{thm:C} lies in
the constraints under which the surfaces appear. The equations are the
BP fixed-point equations of a connected simple cubic graph. Every edge
carries the same coupling. The fixed fields lie in one bounded box that
does not depend on the genus. The two free parameters are vertex fields,
so the surfaces are response components in the sense used throughout the
paper. Under these constraints the number of vertices needed is
determined up to constants, by an algebraic bound on the number of
critical points of a linear height on a response component.

Two readings of Theorem~\ref{thm:C} are not intended. First, the
constructed surfaces consist of saddle points of the Bethe free energy
(Lemma~\ref{lem:negative-index}). The theorem does not produce stable
fixed points of the iteration of arbitrary genus, and the geometry of
the set of all fixed points is not what is being classified. Second,
\(n_{\min}(g)\) counts vertices. The construction does not control,
uniformly in the genus, the precision with which the fixed fields must
be specified or the condition number of the response, and the theorem
makes no claim about them.

Section~\ref{sec:field-map} gives the common coordinates and distinguishes
regularity of the response from regularity of its field projection.
Section~\ref{sec:finiteness} proves finiteness and the multiplicity bounds.
Section~\ref{sec:rigidity} classifies positive-direction sweeps.
Section~\ref{sec:topology} constructs surfaces by iterating a doubling
operation and opens with an overview of the construction.
Section~\ref{sec:uniform} implements that construction with
uniform couplings and bounded fixed fields, and proves the matching
size bound. Section~\ref{sec:diagrams} obtains the topology of the
product example from certified root counts.

\paragraph{Formal verification.}
Theorems~\ref{thm:A}, \ref{thm:B}, and~\ref{thm:C} have been formally
verified in Lean~4 with Mathlib, at the revision \texttt{\leancommit}
of the accompanying repository \cite{responselean}. The verified
statements include the hypotheses and conclusions as stated above, with
genus formalized through a connected-sum presentation whose rational
Betti numbers are \(1,2g,1\). The formal proofs of Theorems~\ref{thm:B}
and~\ref{thm:C} follow routes that differ from the written proofs in
several places, and the quantitative statements of
Sections~\ref{sec:finiteness} and~\ref{sec:diagrams} are not part of the
verified development. Section~\ref{sec:formal} lists the final
declarations, the differences in route, the axiom audit, the
dependencies, and the parts of the paper that remain outside the formal
development.

\section{BP equations and the external-field map}
\label{sec:field-map}

Let \(G=(V,E)\) be finite and undirected, with no self-loops, and let
\(\mathcal A\) consist of the two orientations of each edge. For an arc
\(\alpha=i\to j\) write \(\bar\alpha\) for its reverse on the same edge.
Put \(a_e=e^{2J_e}\), \(b_i=e^{2h_i}\), and let \(y_\alpha\) be the ratio
of the two components of the interaction message sent along \(\alpha\).
The cavity and full vertex ratios are
\[
 X_\alpha=b_i\!\!\!\prod_{\substack{\gamma\to i\\\gamma\ne\bar\alpha}}\!\!\!
       y_\gamma,\qquad
 F_i=b_i\prod_{\gamma\to i}y_\gamma .
\]
With \(M_a(X)=(aX+1)/(X+a)\), the fixed-point equations are
\begin{equation}
 y_\alpha=M_{a_e}(X_\alpha),\qquad
 f_\alpha=y_\alpha(X_\alpha+a_e)-(a_eX_\alpha+1)=0.
 \label{eq:bp}
\end{equation}
An admissible complex root is a zero of \(f\) on the open set where
\begin{equation}
 \left(\prod_\alpha y_\alpha\right)
 \left(\prod_\alpha (X_\alpha+a_{e(\alpha)})\right)
 \left(\prod_i(1+F_i)\right)\ne 0 .
 \label{eq:admissible}
\end{equation}
For the two cavity ratios \(X,Y\) on an edge, the normalizer is
\(aXY+X+Y+a=(1+F_i)(Y+a)\) if \(X\) is the cavity at \(i\).
Here \(F_i=X M_a(Y)\). Thus~\eqref{eq:admissible} also permits normalized
edge beliefs. Positive roots have \(a,b,y>0\) and automatically satisfy
these conditions. Their magnetizations \(m_i=(F_i-1)/(F_i+1)\) can have
either sign.

\subsection{Magnetizations and the reduced Bethe potential}

Write \(u_i=(1+m_i)/2\) and let \(q_e\) be the \(++\) entry of the
edge belief. Its other entries are \(u_i-q_e,u_j-q_e,1-u_i-u_j+q_e\).
For given \(m\) and \(a\) the edge equation is
\begin{equation}
 q_e(1-u_i-u_j+q_e)=a_e^2(u_i-q_e)(u_j-q_e).
 \label{eq:odds}
\end{equation}
The odds ratio on the positive interval
\((\max(0,u_i+u_j-1),\min(u_i,u_j))\) increases strictly from zero to
infinity. Its logarithmic derivative is the sum of the reciprocals of
the four entries. The positive solution is unique, analytic, and
semialgebraic in \((a,m)\), including at \(a_e=1\). From the resulting
edge table, the incoming message at \(i\) is
\[
 y_{j\to i,e}=\frac{F_i(1-u_i-u_j+q_e)}{a_e(u_i-q_e)},\qquad
 F_i=\frac{u_i}{1-u_i}.
\]
The analogous message at \(j\) and \(b_i=F_i/\prod_{\alpha\to i}y_\alpha\)
recover the fixed point. This is an invertible Nash parametrization of
the entire positive family by \((a,m)\), where Nash means real analytic
and semialgebraic. The recovery formulas and the nonzero derivative
in~\eqref{eq:odds} give the same equivalence in local analytic rings
near a positive root, including after complexification.
In particular, fixing \((a,b)\) gives isomorphic local analytic
fixed-point rings in the two presentations. The message Jacobian
and \(D_mH_J\) consequently have the same nullity at a positive
root. This also identifies the local complex lengths used below.

At fixed \(J\), let \(H_{J,i}=\tfrac12\log b_i\) in these coordinates.
Let \(S_i(m_i)\) denote the negative entropy of the binary vertex
belief. Let \(L_e(m_i,m_j)\) denote the minimum of
\(\sum_{\sigma,\tau}p_e(\sigma,\tau)\log p_e(\sigma,\tau)
-J_e\sum_{\sigma,\tau}\sigma\tau p_e(\sigma,\tau)\) at the prescribed
marginals. The strict minimization in \(q_e\) gives~\eqref{eq:odds}.
Equivalently, \(L_e\) is the Legendre transform in the two vertex
fields of the log partition function of one edge. The reduced Bethe
potential is
\begin{equation}
 \mathcal B_J(m)=\sum_i S_i(m_i)
       +\sum_{e=ij}\bigl(L_e(m_i,m_j)-S_i(m_i)-S_j(m_j)\bigr).
 \label{eq:bethe}
\end{equation}
The derivative of \(L_e\) in \(m_i\) is the cavity field
\(\frac12\log X_{i\to j,e}\) of its edge table. The derivative of
\(S_i\) is \(U_i=\atanh m_i\). Since
\(\frac12\log X_{i\to j,e}=U_i-\frac12\log y_{j\to i,e}\), the
gradient of \(\mathcal B_J\) is \(H_J\). Thus \(K=D_mH_J\) is the
Hessian after edge elimination. The Morse index of a fixed point
below is the number of negative eigenvalues of \(K\). The edge
directions eliminated from the Bethe potential are positive
definite, so they do not change that index. \cite{yedidia,zeta}

Let \(v_i=1-m_i^2\) and let
\[
 \beta_e=\frac{\mathbb E_e[\sigma_i\sigma_j]-m_im_j}
                   {\sqrt{v_iv_j}}\in(-1,1)
\]
be the normalized covariance under the positive edge belief.
The Hessian of \(L_e\) is the
inverse covariance matrix of these spins. Thus
\(A=\diag(\sqrt{v_i})K\diag(\sqrt{v_i})\) has entries
\begin{equation}
 A_{ii}=1+\sum_{e\ni i}\frac{\beta_e^2}{1-\beta_e^2},\qquad
 A_{ij}=-\sum_{e=ij}\frac{\beta_e}{1-\beta_e^2}\quad(i\ne j).
 \label{eq:hessian}
\end{equation}
The sign of \(\beta_e\) is the sign of \(J_e\), by~\eqref{eq:odds}.
For a subset \(W\subset V\), the principal block \(A_{WW}\) equals
the normalized Hessian from edges of \(G[W]\), with the same
correlation weights, plus a nonnegative diagonal contribution
from the edges leaving \(W\). In particular, \(K_{WW}\) is positive
definite if \(G[W]\) is a pseudoforest (each component has at most
one cycle). To see this, let \(B(w)_{\alpha\gamma}=w_{e(\alpha)}\)
when \(\gamma\) enters the tail of \(\alpha\) and
\(\gamma\ne\bar\alpha\), and zero otherwise. The weighted
Ihara--Bass identity on the induced graph reads
\[
 \det A_{G[W]}(w)=\frac{\det(I-B(w))}{\prod_{e\in E(G[W])}(1-w_e^2)}.
\]
The numerator is \(1\) on a tree and
\((1-\prod_{e\in C}w_e)^2\) on a unicyclic component with cycle
\(C\). Hence the determinant is positive along
\(w=\tau\beta\), \(0\leq\tau\leq1\). Since the matrix starts
at the identity, no eigenvalue can cross zero. Adding the boundary
diagonal and undoing the positive normalization preserve positive
definiteness. This includes signed weights and a two-edge cycle
formed by parallel edges.
\cite{zeta,signed}

\subsection{Properness, degree, and response components}

Put \(z_\alpha=\tfrac12\log y_\alpha\). In these coordinates the
edge update is \(g_t(x)=\atanh(t\tanh x)\) at \(t=\tanh J_e\), applied
to the cavity field \(x=h_i+\sum_{\gamma\to i,\ \gamma\ne\bar\alpha}
z_\gamma\). In particular, \(\abs{z_\alpha}\leq\abs{J_e}\) and
\begin{equation}
 \abs{H_{J,i}(m)-\atanh m_i}\leq\sum_{e\ni i}\abs{J_e}.
 \label{eq:proper}
\end{equation}
This holds for every \(m\) by reconstruction. Thus \(H_J\) is proper.
In the coordinates \(U=\atanh m\), a straight homotopy from \(H_J\) to
\(U\) remains proper by~\eqref{eq:proper}, giving Brouwer degree \(1\).
A nondegenerate positive root contributes
\(\operatorname{sgn}\det K=(-1)^{\operatorname{ind}K}\).
All fields sufficiently far along either end of a positive-direction
sweep have a unique, nondegenerate root. Indeed, the message box
\(\abs{z_\alpha}\leq\abs{J_e}\) is invariant and \(g'_t\) tends
uniformly to zero there as every cavity field tends to its corresponding
infinity.

For an affine field slice \(L\), the response in magnetization
coordinates is \(H_J^{-1}(L)\). In a coordinate slice with free set \(D\)
and \(W=V\setminus D\), the defining equations and projection are
\[
 H_{J,W}(m)=h_W,\qquad \pi(m)=H_{J,D}(m).
\]
A regular component is a connected component on which \(D_mH_{J,W}\)
has full row rank. It is a \(\lvert D\rvert\)-dimensional manifold
whose normal bundle is framed by these defining equations, so it is
orientable. On it a critical point of \(\pi\) is exactly a point where
\(K\) is singular. Regularity of the component allows these critical
points. A fold is a simple quadratic turning of a one-dimensional
projection, or its product with an identity in transverse parameters.
We count roots on a compact component in its ordinary finite-field
domain. The coordinate \(b_D=e^{2h_D}\) makes the response and its
projection semialgebraic, and changes neither root counts nor topology.

A regular compact component contributes zero to the degree of the
full field map. Indeed, choose a small product neighborhood in which
the normal coordinates are \(H_{J,W}-h_W\). Such a neighborhood is
obtained by solving in the framed normal bundle. Its boundary
avoids every value \((h_W,h_D)\), so its local degree is constant
as \(h_D\) varies. The image of the closed neighborhood is bounded,
which makes that degree zero. The global degree \(1\) consequently
requires a positive root outside this neighborhood at each field
in the slice.

\section{Universal finiteness and multiplicity bounds}
\label{sec:finiteness}

\subsection{A rational function stationary at fixed points}

First remove every edge with \(J_e=0\). On such an edge admissibility
forces \(y_\alpha=1\) in both directions. Reinserting those messages
after solving on the support graph preserves admissibility, since
the reinserted update denominator at \(i\) is \(1+F_i\). In the
admissible local rings the residual
\((X_\alpha+1)(y_\alpha-1)\) eliminates \(y_\alpha\) as well.
We can therefore work on the support graph, with \(a_e>0\),
\(a_e\ne1\), and \(b_i>0\).

On \(e=ij\) write \(a=a_e\), \(L=y_{i\to j,e}\), \(R=y_{j\to i,e}\).
The edge equations imply
\begin{equation}
 L=\frac{aF_i+R}{F_i+aR},\qquad
 F_i=R\frac{aL-1}{a-L},\qquad
 F_j=L\frac{aR-1}{a-R}.
 \label{eq:edge-inverse}
\end{equation}
An output of \(M_a\) at a finite allowed input cannot equal \(a\)
when \(a^2\ne1\). Therefore \(a-L\) and \(a-R\) are nonzero.
The edge factor \(Q_e=aLR-L-R+a\) satisfies
\begin{equation}
 Q_e=(1+F_i)(a-L)=(1+F_j)(a-R)\ne0.
 \label{eq:edge-factor}
\end{equation}
Define the nonzero rational function
\[
 \mathcal R(y)=\frac{\prod_i(1+F_i(y))}{\prod_e Q_e(y)}
\]
on the open set where the factors are nonzero. Holding the model
parameters fixed, the logarithmic derivative in the message \(L\) is
\[
 \frac{L}{\mathcal R}\frac{\partial\mathcal R}{\partial L}
 =\frac{F_j}{1+F_j}-\frac{L(aR-1)}{Q_e}.
\]
The two terms agree at every admissible root by
\eqref{eq:edge-inverse} and~\eqref{eq:edge-factor}. Thus all message
derivatives vanish at a fixed point. On an irreducible algebraic curve
of admissible roots, the differential of \(\mathcal R\) vanishes by the
chain rule. Over \(\C\) this makes \(\mathcal R\) a nonzero constant on
the curve. All logarithmic derivatives here are rational differentials.

\subsection{A pole-order budget}

An infinite admissible fiber would have a positive-dimensional
component, since it is an open subset of an affine algebraic set.
One could take an irreducible curve meeting that open set and pass
to the smooth projective model of its function field. On this model
some magnetization is nonconstant. Indeed, fixing all \(F_i\) leaves
at most two choices on each edge: elimination in~\eqref{eq:edge-inverse}
gives
\begin{equation}
 a(1+F_j)R^2+
 \bigl((a^2+F_j)F_i-(1+a^2F_j)\bigr)R-aF_i(1+F_j)=0 .
 \label{eq:edge-quadratic}
\end{equation}
Its leading and constant coefficients are nonzero at an admissible
point, and \(L\) is recovered from \(R\). Since \(m_i\) determines
\(F_i\), the fiber of the magnetization tuple on the admissible
locus is finite.

Fix a pole \(P\) of some \(m_i\) on the smooth projective curve.
Let \(\delta_i\) be the pole order of \(m_i\) there, taken as zero
when \(m_i\) is regular, including when it is identically zero.
We can apply a single spin change to the entire curve so that all
\(F_i\) are regular at \(P\). Namely, choose \(\epsilon_i=-1\) when
\(F_i\) has a pole there and \(+1\) otherwise. The transformations are
\[
 a_{ij}\mapsto a_{ij}^{\epsilon_i\epsilon_j},\qquad
 b_i\mapsto b_i^{\epsilon_i},\qquad
 y_{i\to j,e}\mapsto y_{i\to j,e}^{\epsilon_j}.
\]
They give the admissible family for a new fixed real model, by
\(M_a(1/X)=1/M_a(X)=M_{1/a}(X)\). All \(a_e,b_i\) remain positive,
and \(F_i\mapsto F_i^{\epsilon_i}\), \(m_i\mapsto\epsilon_i m_i\).
In these coordinates,
\[
 \delta_i=\ord_P(1+F_i).
\]
At least one of \(L,R\) is regular at \(P\) on each edge. If both had
poles, \(F_i=R(aL-1)/(a-L)\) would have a pole since the quotient
tends to \(-a\). By~\eqref{eq:edge-factor}, choosing the regular
message gives
\[
 \ord_P Q_e\geq\min(\delta_i,\delta_j).
\]
Stationarity applies to \(\mathcal R\) in the transformed model.
Taking its valuation gives the budget
\begin{equation}
 \sum_i\delta_i=\sum_e\ord_P Q_e
      \ \geq\ \sum_{e=ij}\min(\delta_i,\delta_j).
 \label{eq:pole-budget}
\end{equation}

\subsection{Every positive pole layer contains excess edges}

For an edge \(e=ij\), the identities above imply
\begin{equation}
 1+F_j=k_{ij}(1+F_i),\qquad
 k_{ij}=\frac{(a^2-1)R}{(F_i+aR)(a-R)}.
 \label{eq:edge-transport}
\end{equation}
If \(\delta_i>\delta_j\), this coefficient has negative order at
\(P\), and \(F_i(P)=-1\). If \(R\) vanishes or has a pole, the
order of \(k_{ij}\) is positive. At a finite nonzero \(R(P)\) the
denominator can vanish only for \(R(P)=a\) or \(a^{-1}\).
Thus the message into the higher-order vertex has strictly positive
value at \(P\).

For \(r\geq1\), let \(S_r=\{i:\delta_i\geq r\}\), and let \(T\)
be a nonempty connected component of \(G[S_r]\). Every incoming
message from outside \(T\) is a unit at \(P\) with positive value.
Absorb those messages into the vertex field ratios on \(T\). At
each vertex the resulting external factor is a unit with value
\(\eta_i>0\), and \(F_i(P)=-1\).

Suppose first that \(T\) is a tree. At a leaf \(i\) of \(T\) the
only internal input must have value \(R(P)=-1/\eta_i<0\).
Its reverse has value \((R(P)-a)/(aR(P)-1)>0\) by the edge
equation. It is a unit and can be absorbed into the factor of the
neighbor when the leaf is removed. This leaves the same conditions
on the smaller graph. An isolated vertex has \(F_i(P)=\eta_i>0\),
a contradiction. This reasoning only evaluates internal messages
after identifying the other inputs as units.

If \(T\) is unicyclic, remove leaves in the same manner. In one
cyclic orientation the message values \(x_i\) on the remaining cycle
satisfy
\[
 x_i=M_{a_i}(\eta_i x_{i-1}).
\]
The update is defined over the local ring as a projective linear
transformation: its determinant \(\eta_i(a_i^2-1)\) at \(P\) is
nonzero. Thus the recurrence can first be evaluated in
\(\mathbb P^1(\C)\), even if a message has a pole. The product
around the cycle has a strictly positive real matrix
\(\bigl(\begin{smallmatrix}A&B\\C&D\end{smallmatrix}\bigr)\).
Its fixed-point equation \(Cx^2+(D-A)x-B=0\) has one positive and
one negative real root. Every cyclic message is finite and
nonzero, by starting the product at each edge. A positive cyclic
message makes the others positive, so the values have a common
sign. If the cyclic values are negative, the reverse values are
positive by \(F_i(P)=-1\). Choose the orientation in which the
cyclic values are positive.

Now \(U_i=\eta_i x_{i-1}>0\) and the reverse input on \(i\to i+1\)
is \(-1/U_i\). Substitution into~\eqref{eq:edge-transport} gives
\[
 \abs{k_{i,i+1}(P)}
 =\frac{\abs{a_i^2-1}U_i}{(U_i+a_i)(a_i U_i+1)}<1.
\]
In the function field the product of the \(k_{i,i+1}\) around the
cycle is \(1\), since the \(1+F_i\) telescope. The displayed
strict inequality contradicts its evaluation at \(P\).
The cycle argument includes two distinct parallel edges.
Thus every component of every nonempty layer has cycle rank at
least two.

Let \(c_r\) count the components of \(G[S_r]\).
Then \(\lvert E(G[S_r])\rvert\geq\lvert S_r\rvert+c_r\).
Summing the layer counts yields
\[
 \sum_{e=ij}\min(\delta_i,\delta_j)
 =\sum_{r\geq1}\lvert E(G[S_r])\rvert
 \geq\sum_i\delta_i+\sum_{r\geq1}c_r
 >\sum_i\delta_i.
\]
The last inequality uses the choice of a pole. This contradicts
\eqref{eq:pole-budget}, proving Theorem~\ref{thm:A}. The positivity
of the real model coefficients is used at the layer boundary and
in the cycle argument.

\subsection{Multiplicity and the index of a degenerate root}

For an admissible root \(x\), let \(\ell_x\) be the complex length
of its local fixed-field ring. For \(A\subseteq E(G_0)\) write
\(d_A(i)\) for its degree at \(i\), and let
\[
 Q(G_0)=\sum_{A\subseteq E(G_0)}\prod_{i\in V}\phi(d_A(i)),
 \qquad \phi(0)=1,\qquad \phi(d)=d-1\quad(d\geq1).
\]

\begin{corollary}[Special-fiber multiplicity]
\label{cor:budget}
Each admissible fixed-field root is an isolated local complete
intersection. For any fixed finite real model,
\[
 \sum_{x\in X^{\rm adm}_{J,h}}\ell_x\leq Q(G_0),\qquad
 \sum_{x\in X^+_{J,h}}\ell_x\leq Q(G_0)-2^{\nu(G_0)}+1 .
\]
In each Eulerian real message-sign sector the sum of lengths of
real roots is a positive odd integer.
\end{corollary}

\begin{proof}
Work on \(G_0\) by regular elimination of the zero edges.
In the message coordinates the isolated fiber is defined by
\(2\lvert E(G_0)\rvert\) equations in the same number of variables.
These equations form a system of parameters and hence a regular
sequence in the smooth ambient local ring. The full admissible
family is smooth of dimension \(\lvert E(G_0)\rvert+\lvert V\rvert\),
equal to that of its coefficient space. One can see this in
coordinates \((u_i,q_e,a_e)\): the edge equations are
\eqref{eq:odds} with all four entries units and \(a_e\ne0\),
and their \(a_e\) derivatives are
\(-2a_e(u_i-q_e)(u_j-q_e)\ne0\), independently for each edge.
These coordinates and the message formulas of
Section~\ref{sec:field-map} are inverse on the admissible locus.
The projection is finite as an analytic germ at each isolated
fiber point, by isolating the point in a small ball and keeping the
boundary away from the base parameter. The source is
Cohen--Macaulay, the target regular, and their dimensions agree.
Miracle flatness gives local freeness of rank \(\ell_x\).
The analytic and algebraic lengths agree for a finite local ring.
\cite{counting,stacks}

Choose disjoint neighborhoods for all roots in the fiber.
A nearby point of the generic finite \'etale locus then has
\(\ell_x\) simple roots in the neighborhood of \(x\). Since its
total is \(Q(G_0)\), this proves the first bound. The nearby
parameter can also be chosen real, in the same nonzero coupling
sign chamber, since the complement of the complex Zariski open
has empty interior on the real parameter space. \cite{counting}

For the real count, the Eulerian sector \(S\) is indexed by the
set of edges whose two messages are negative. There are
\(2^{\nu(G_0)}\) such sectors. The real sector coordinates of
\cite{counting} use \(U_i=\frac12\log F_i\). For fixed \(U\), the
two edge branches of~\eqref{eq:edge-quadratic} have opposite
signs, with product of their incoming ratios \(-F_i\).
Indeed, for \(F_i,F_j>0\) the leading coefficient is positive
and the constant coefficient is negative, so the discriminant
is strictly positive. Both branches and their recovered reverse
messages are locally analytic, and each pair has the same sign.
Choosing the negative pair on \(S\) gives positive field ratios
precisely when every \(d_S(i)\) is even. Thus each Eulerian
choice defines a sector parametrized by all \(U\in\R^V\).
Consequently the external-field map on sector \(S\) has the form
\[
 (\mathfrak h_S)_i(U)=(1-d_S(i))U_i-
            \sum_{e\ni i}\varepsilon_e z^+_{e,i}(U),
 \qquad
 \varepsilon_e=\begin{cases}-1,&e\in S,\\1,&e\notin S,\end{cases}
\]
where \(z^+_{e,i}\) is the half-log positive incoming message
and is bounded by \(\abs{J_e}\). Each \(d_S(i)\) is even. The
proper degree is therefore that of the invertible diagonal
linear part, \((-1)^{v(S)}\), with \(v(S)\) the number of vertices
incident to \(S\). In particular, every sector contributes at
least one real root. Subtracting the nonempty sectors from the
complex length bound gives the positive-root bound.

For a real root \(x\) in sector \(S\), take its local Brouwer
degree \(d_x^{\rm vert}\) in these vertex coordinates. Under a
small generic real perturbation as above, the real simple roots
near \(x\) have indices \(\pm1\), and the nonreal roots occur in
conjugate pairs. Additivity of the local degree yields
\[
 |d_x^{\rm vert}|\leq\ell_x,\qquad
 d_x^{\rm vert}\equiv\ell_x\pmod2,\qquad
 \sum_{x\ \text{in }S}d_x^{\rm vert}=(-1)^{v(S)}.
\]
The length sum in \(S\) is therefore positive and odd.
\end{proof}

The orientation can also be expressed using the fixed-point residual
\(y-M(X(y))\). Its local degree is \(d_x^{\rm msg}=(-1)^{|S|}d_x^{\rm vert}\),
so its signed sector sum is \((-1)^{|S|-v(S)}\). To see the
orientation factor at a simple root, adjoin the vertex consistency
equations for independent \(F\). With those \(F\) held fixed, the
two cleared residuals on an edge are
\[
 L(F_i+aR)-(aF_i+R),\qquad R(F_j+aL)-(aF_j+L).
\]
Eliminating \(L\) in the second residual gives the quadratic
in~\eqref{eq:edge-quadratic} divided by \(F_i+aR\).
Thus the edge Jacobian determinant of the cleared pair is the
quadratic's derivative at \(R\). It is positive on the positive
branch and negative on the negative branch. The product of the
two clearing factors is positive on either branch, since on the
negative branch each cavity ratio lies strictly between
\(-a\) and \(-a^{-1}\). Hence clearing does not change the edge
orientation. To check the vertex factor, use consistency residuals
\(G_i=F_i-b_i\prod_{\alpha\to i}y_\alpha(F)\).
At a root their derivative is
\[
 D_FG=\diag(F)\,D_U\mathfrak h_S\,\diag(F)^{-1},
\]
since the reconstructed field ratio is
\(\exp(2(\mathfrak h_S)_i)\). Thus the vertex block has the
orientation of \(\mathfrak h_S\). Eliminating the independent
\(F\) first has identity consistency block and recovers the
original message residual. Eliminating the edge variables first
therefore supplies exactly the factor \((-1)^{|S|}\).
The simple-root calculation extends to degenerate roots by
additivity under the same real perturbation.

\section{Graph-theoretic rigidity of one-parameter response}
\label{sec:rigidity}

In this section all couplings are finite and strictly positive.
We first prove the order property used to rule out circular components.
We then show how a pair of folds creates a circle on every graph
with \(p_U\geq2\).

\subsection{Order and the lowest Hessian eigenvalue}

\begin{lemma}[Sign domains]
\label{lem:sign-domains}
Suppose \(A_0\) is a symmetric matrix with strictly negative
off-diagonal entries on the edges of \(G\), zero entries at nonedges,
and positive definite principal blocks on all induced pseudoforests.
For any eigenvector of \(A_0\) of eigenvalue \(\lambda\leq0\), each
connected component of its positive or negative support has cycle
rank at least two.
\end{lemma}
\begin{proof}
Let \(I\) be a connected component of the positive support of \(x\).
All entries of \(x\) adjacent to \(I\) outside \(I\) are nonpositive.
Thus
\[
 (A_0)_{II}x_I=\lambda x_I-(A_0)_{I,V\setminus I}x_{V\setminus I}\leq 0.
\]
The quadratic form on \(x_I>0\) cannot be positive. The same
argument for \(-x\) handles the negative support.
\end{proof}

\begin{proposition}[Coordinatewise order of roots]
\label{prop:order}
Let \(p_U(G)\leq1\) and let all couplings be positive. Any two
positive roots at the same field are comparable in the coordinatewise
order. If \(G\) is connected, two distinct roots are strictly ordered
in every coordinate.
\end{proposition}

\begin{proof}
If \(m,m'\) are two positive roots at the same field, their difference
is in the kernel of
\(\overline K=\int_0^1K(m'+t(m-m'))\,dt\), by the fundamental
theorem of calculus. Each pseudoforest principal block of
\(\overline K\) is positive definite because it is an average of
positive definite blocks. Its edge entries are strictly negative
by~\eqref{eq:hessian}. Thus it meets the hypotheses of
Lemma~\ref{lem:sign-domains}.
For \(p_U\leq1\) the difference has one sign. On a connected graph,
a zero entry then forces all its neighboring entries to vanish by
the kernel equation, and hence all entries to vanish. Distinct
roots are therefore strictly ordered in every coordinate.
For a disconnected graph with \(p_U\leq1\) there is at most one
component that is not a pseudoforest. On the other components
strict convexity and proper degree one give a unique nondegenerate
root at each field. Consequently the full set of roots is ordered
coordinatewise.
\end{proof}

For a connected graph the smallest eigenvalue of \(K\) is simple
with a strictly positive eigenvector, by Perron--Frobenius applied
to \(cI-K\) for large \(c\). Any other eigenvector is orthogonal to
it and has both signs. If \(p_U\leq1\), the sign-domain lemma gives
strict positivity of every eigenvalue except possibly the smallest.
In particular at a singular point
\(\ker K=\R q\) for \(q>0\), and all other eigenvalues are positive.

\subsection{A single line when \texorpdfstring{\(p_U\leq1\)}{pU is at most one}}

Take \(G\) connected, \(p_U\leq1\), and a sweep \(h=h^0+sv\) with
\(v>0\). The defining differential \([K,-v]\) has full row rank
because at a singular point \(q^\top v>0\). The full response
is a real analytic curve, oriented by the nowhere vanishing tangent
\begin{equation}
 T=(\adj(K)v,\det K).
 \label{eq:tangent}
\end{equation}
At a critical point of the projection to \(s\),
\(\adj(K)=cqq^\top\) for \(c>0\), so every magnetization
coordinate of the tangent is positive.

Fix a vertex \(i\). Projection to \((m_i,s)\) is injective by the
strict order at fixed \(s\). It is an immersion, since at a turning
point the \(m_i\) derivative is positive and elsewhere the \(s\)
derivative is nonzero. It is proper by~\eqref{eq:proper} and
therefore embeds the curve in the plane. A circle component
would give an oriented simple closed curve with all its horizontal
tangents pointing to the right. At points on its upper and lower
horizontal supporting lines the interior lies on opposite sides
of the horizontal line, so the two oriented tangents must point in
opposite directions. Thus the response has no circle component.
This argument allows higher-order turning points.

Each remaining component is a line. Properness of the projection to
\(s\) forces each end to tend to one of \(+\infty,-\infty\). If the
two ends tend to the same infinity, sufficiently large fields on that
side have at least two roots, contrary to the strong-field contraction
in Section~\ref{sec:field-map}. Hence each component joins the two
infinities, and uniqueness at the ends permits just one component.
The critical points are confined to a compact portion of this
analytic line, where \(\det K\) has only finitely many zeros because
it is nonzero near the ends. For a disconnected graph the unique
analytic roots on the pseudoforest components can be adjoined to the
response on the one remaining component. If all components are
pseudoforests the full response is a graph over \(s\).
This proves the forward implication of Theorem~\ref{thm:B}.

\subsection{Two folds produce a circle}

We use the first ferromagnetic bifurcation of a connected subgraph
whose cycle rank is at least two. Its leaf-pruned core has minimum
degree two and is not a cycle. On that core let \(B_{\rm nb}\)
be the nonbacktracking matrix, with rows indexed by output arcs
and columns by allowed input arcs. It is irreducible. In fact,
after deleting the edge of an arriving arc there is a cycle
reachable from its head in the remaining component. For a bridge
this follows from minimum degree two on either side except
possibly at the attachment vertex. Going to that cycle, around it,
and back allows a nonbacktracking walk from the arriving arc to its
reverse. Connectivity now gives irreducibility. Since every row
has at least one entry and some row has more than one,
\(\rho(B_{\rm nb})>1\).

Put the common value \(\tanh J=\tau_*=1/\rho(B_{\rm nb})\)
at the critical point. Write \(r,l>0\) for right and left Perron
vectors of \(\tau_*B_{\rm nb}\), normalized by \(l^\top r=1\).
Use \(\tanh J=(1+\mu)\tau_*\) and common external field \(\eta\).
In log-message coordinates write \(z=xr+w\), \(l^\top w=0\).
The projection of the BP residual to the complementary space can
be solved for \(w\) by the implicit function theorem, because the
eigenvalue \(1\) is simple. At \(\eta=0\) this solution is odd and
\(w=O(x^3)\). From
\(g_t(u)=tu-t(1-t^2)u^3/3+O(u^5)\) the remaining equation is
\begin{equation}
 -\mu x+\kappa x^3-\gamma\eta+
 O(x^5+|\mu|\,|x|^3)
 +O\bigl(|\eta|(|x|+|\mu|+|\eta|)\bigr)=0,
 \label{eq:perron-cubic}
\end{equation}
where
\[
 \kappa=\frac{1-\tau_*^2}{3\tau_*^2}\sum_\alpha l_\alpha r_\alpha^3>0,
 \qquad
 \gamma=\tau_*\sum_\alpha l_\alpha>0 .
\]
The field derivative is nonzero. With \(x=\sqrt{\mu}\,X\) and
\(\eta=\mu^{3/2}Y\) for small \(\mu>0\), the field graph in these
coordinates converges in \(C^2\) on compact intervals to
\(Y=(\kappa X^3-X)/\gamma\). There is a nondegenerate maximum at
positive field and a nondegenerate minimum at negative field.
Both folds lie on smooth portions of the full response.

Suppose \(p_U\geq2\) and choose vertex-disjoint cores \(G_1,G_2\).
Temporarily set all other couplings to zero. Near the maximum of
the first core and the minimum of the second, let their field
graphs be \(\varphi_1(x_1),\varphi_2(x_2)\).
The difference has a nondegenerate strict maximum \(M\) at the
pair of folds. Choose a small \(\delta>0\). The level
\(\varphi_1(x_1)-\varphi_2(x_2)=M-\delta\) in a small neighborhood is
a smooth circle by the Morse lemma. It is exactly the local
response for the full-graph sweep
\[
 h_i(s)=\begin{cases}
 s-(M-\delta),&i\in V(G_2),\\
 s,&i\notin V(G_2).
 \end{cases}
\]
All unused vertices are isolated at this decoupled setting and
have the unique roots \(m_i=\tanh s\). The full response equations
have full row rank along
the circle since the difference of the two field graphs has
nonzero differential there. The circle lies inside the local
coordinate neighborhood, and near each of its points the entire
zero set is this one-dimensional manifold. It is consequently
open in the full zero set and, by compactness, closed there.
It is thus a connected component of the full response.

\begin{lemma}[Persistence of a regular compact component]
\label{lem:persistence}
Let \(F\colon\Omega\subset\R^N\to\R^r\) be smooth and let
\(C\) be a compact connected component of \(F^{-1}(0)\) on which
\(DF\) has rank \(r\). There is a relatively compact tubular
neighborhood \(U\) of \(C\) such that every smooth map sufficiently
close to \(F\) in \(C^1(\overline U)\) has exactly one zero-set
component in \(U\). This component is regular, diffeomorphic to
\(C\), and is a connected component of the full zero set.
\end{lemma}
\begin{proof}
On the normal bundle of \(C\), \(DF\) is an isomorphism to
\(C\times\R^r\), with right inverse
\(DF^\top(DF\,DF^\top)^{-1}\). The parameter-dependent implicit
function theorem in these normal coordinates produces a unique
small normal graph of zeros. Compactness makes the estimates
uniform. Shrink \(U\) so that the original zero set in
\(\overline U\) is precisely \(C\); this is possible by local
regularity and compactness. On the remaining closed annulus
\(\abs F\) has a positive minimum, which excludes other zeros
and boundary zeros under a small perturbation. The new graph
is compact and open in the full zero set, so it is a connected
component. For \(C^q\)-small perturbations the graph is
\(C^q\)-close, for any fixed finite \(q\geq1\).
\end{proof}

Applying Lemma~\ref{lem:persistence} when
the unused couplings are made small and strictly positive gives
a regular circle on \(G\) with a single common field direction.
The parameter choice witnesses failure of the universal
line property. This proves the converse of Theorem~\ref{thm:B}.

\paragraph{Bethe-potential route to the converse.}
The formal proof obtains the two turning arcs from three zero-field
roots on each connected core of cycle rank two. With a sufficiently
large common \(a=\exp(2J)\), the zero-magnetization Hessian on such a
core satisfies
\[
  \ones^{\mathsf T} K(0)\ones=|V|-|E|(1-a^{-1})<0.
\]
The reduced Bethe potential has an interior minimum. Indeed, the
bounded message contributions and the diverging \(\operatorname{atanh} m_i\)
term in \eqref{eq:proper} ensure that clamping the coordinates to a
sufficiently large closed box inside the open cube cannot increase the
potential. The origin is not a minimum by the Hessian inequality,
and evenness gives a pair of opposite nonzero minimizers. Together
with the origin these give three zero-field roots.

Each core of rank two satisfies \(p_U\leq1\). Its analytic response
line has opposite infinite limits of the sweep height and takes the
height zero at least three times. It therefore has a strict local
maximum and a strict local minimum, and these have isolated critical
times. In small arcs about the selected extrema, the difference of
heights has the form \(M-f_1(x_1)-f_2(x_2)\), with both \(f_i\)
zero at their centers, positive elsewhere and strictly monotone on
each side. The signed square roots of the \(f_i\) give continuous
local coordinates identifying a nearby level with a circle. On this
level at least one of the height derivatives is nonzero, giving full
row rank for the joint sweep equations. Thus the circle is a smooth
compact connected component, and Lemma~\ref{lem:persistence} applies.
This construction requires regularity on that component. The full
response at the same parameters can contain singular points.

\subsection{Forbidden minors and the zero-field classification}

Every connected simple graph of cycle rank at least two
has a connected subgraph of cycle rank two. Pruning its leaves
leaves either a theta graph, two cycles sharing a vertex, or two
cycles joined by a path. In a simple theta graph at most one of
the three paths has length one. Contraction gives a diamond
with path lengths \(1,2,2\). In the other two cases contraction
gives the bowtie. Two vertex-disjoint subgraphs of cycle rank at
least two therefore give one of \(B\sqcup B\), \(B\sqcup D\),
\(D\sqcup D\) as a minor. Conversely, the two sides of such a
minor model lift to disjoint connected subgraphs of rank at
least two, since contraction and deletion cannot increase cycle
rank. Every proper simple minor of \(B\) or \(D\) is a pseudoforest,
which gives minimality.

For a connected graph with \(p_U=1\), there is a sharp zero-field
threshold. Let \(L=\diag(\tanh J_{e(\alpha)})B_{\rm nb}\) on its
leaf-pruned core. At \(h=0\) every leaf-to-core message vanishes
in log coordinates, and the remaining tree messages are uniquely
reconstructed. If \(\rho(L)\leq1\), the strict secant inequality
\(\abs{g_t(u)-g_t(w)}<t\abs{u-w}\) for \(u\ne w\), summed
against the positive left Perron vector, implies uniqueness.
The root is zero. It is simple for \(\rho(L)<1\) and has
corank one at equality.

If \(\rho(L)>1\), iteration from a sufficiently small multiple of the
positive Perron vector increases to a positive core fixed point in the
message box. Strict concavity of \(g_t\) for positive inputs gives
uniqueness in the positive cone: if two strictly positive fixed
points had maximal coordinate ratio \(\lambda>1\), then
\(T(z)\leq T(\lambda z')<\lambda T(z')\) at every coordinate.
At the positive solution, \(D T(z)z<z\) also implies simplicity.
Odd symmetry gives the corresponding negative log-message solution.

To exclude other roots, compare their magnetizations to zero using
the strict order. If the magnetizations are positive then all log
messages are nonnegative. Indeed, write
\(U_i=h_i+P_i-N_i\), with \(P_i,N_i\) the sums of positive
inputs and of absolute values of negative inputs. For \(h_i,U_i\geq0\),
a negative output must use a reverse input \(p>U_i\) and has magnitude
strictly less than \(p-U_i\). The sum of magnitudes of negative
outputs at \(i\) is at most \(N_i\), strictly less whenever there is
a negative output, since \(\sum_{p>U_i}(p-U_i)\leq P_i-U_i\)
if that sum is nonempty. Summing over vertices equates input and
output totals, so there are no negative outputs. This argument
also works for linear updates \(z_{i\to j}=t_e(U_i-z_{j\to i})\)
with \(0<t_e<1\). Irreducibility propagates a nonzero nonnegative
core fixed point to strict positivity. The roots below zero
follow by spin reversal.

The central root for \(\rho(L)>1\) is simple as well. If \(K(0)\)
were singular it would have a strictly positive kernel vector \(q\).
Writing \(t_e=\tanh J_e\), the entries
\[
 R_{i\to j}=\frac{t_e(q_i-t_e q_j)}{1-t_e^2}
\]
satisfy \(\sum_{j\to i}R_{j\to i}=q_i\) by~\eqref{eq:hessian},
and \(R_{i\to j}=t_e(q_i-R_{j\to i})\). The linear sign argument
makes \(R\geq0\). The core part is nonzero, since otherwise the
tree reconstruction would give \(R=0\) and \(q=0\).
Thus the irreducible core has \(LR=R\) with \(R\geq0\) nonzero,
contrary to \(\rho(L)>1\).

\begin{proposition}[Zero-field classification]
\label{prop:zero-field}
Over positive couplings at which all zero-field roots are
nondegenerate, the maximum number of zero-field roots is \(1\) when
\(p_U(G)=0\), \(3\) when \(p_U(G)=1\), and at least \(9\) when
\(p_U(G)\geq2\). The universal coordinatewise order of the roots at
every field, and the universal bound of three distinct positive roots
at zero field, are each equivalent to \(p_U(G)\leq1\).
\end{proposition}

\begin{proof}
The first two cases follow from the threshold analysis above, applied
to the unique non-pseudoforest component, together with the unique
roots on the pseudoforest components. For the last case
choose two disjoint connected rank-two subgraphs. At decoupling,
each can be made strictly supercritical, with three simple roots.
Together with the isolated unused vertices this gives exactly
nine simple roots at decoupling. The implicit function theorem
preserves these roots when the unused couplings are turned on.
There are no other roots for sufficiently small added couplings:
\eqref{eq:proper}, uniformly on a compact parameter neighborhood,
would make any sequence of additional roots converge to one of
the nine roots, contradicting its local uniqueness. Thus the
perturbed zero-field model still has all roots nondegenerate.
The two choices of opposite magnetic signs on the cores also give
a pair of incomparable roots. Thus the universal coordinatewise
order at all fields and the universal bound of three distinct
positive roots at zero field both fail when \(p_U\geq2\), and both
hold when \(p_U\leq1\) by Proposition~\ref{prop:order} and the
threshold analysis.
\end{proof}

\subsection{One state, with arbitrarily many crossings}

For \(p_U\leq1\) the uniform sweep has a Nash one-state representation
that works simultaneously for all positive couplings and all field
ratios. Fix a reference vertex \(n\) and put \(r_i=b_i/b_n\) for
\(i\ne n\).

\begin{proposition}[Nash one-state representation]
\label{prop:nash}
Let \(p_U(G)\leq1\). Over
\((a,r)\in(1,\infty)^E\times(0,\infty)^{n-1}\),
the entire positive family is Nash-isomorphic to the product of
this base with \((0,1)\), preserving \((a,r)\). In these coordinates
the field projection is a scalar Nash function
\(b_n=\mathcal B(a,r,s)\) with limits \(0,\infty\) at the two ends
of the state interval.
\end{proposition}

The proof uses a compactification by the
two strong-field ends and is given in Appendix~\ref{app:nash}.

A large number of roots can still lie on one curve. For any integer
\(N\geq1\), consider a cactus formed from a root triangle and \(N\) separate leaf triangles,
each joined by one bridge to the same root vertex. Deleting this
vertex leaves a pseudoforest, and the graph has \(p_U=1\).
For a zero-field triangle away from its attachment vertex, let
\(d=(1-\prod_{e\ {\rm on\ cycle}}\tanh J_e)/
(1+\prod_{e\ {\rm on\ cycle}}\tanh J_e)\).
At prescribed full field \(U\) at the attachment vertex, the two
return messages are equal by strict contraction around the cycle.
Their sum is \(U-A_d(U)\), where
\(A_d(U)=\asinh(d\sinh U)\). This follows by substituting
\((U-A_d(U))/2\) into the return-message equation.

For a leaf triangle use parameter \(d=\epsilon\). If the bridge has
\(\tanh J=k\) and the leaf attachment field is \(j\), its return
message \(p\) is the unique solution in \((-J,J)\) of
\[
 p=g_k\bigl(j+R_\epsilon(j+g_k(U-p))\bigr),\qquad
 R_\epsilon(x)=\asinh(\epsilon^{-1}\sinh x)-x .
\]
In fact \(R_\epsilon'>0\), so the right-hand side decreases in \(p\).
Take \(j=-g_k(c)\) for any \(c\). As \(\epsilon\downarrow0\),
\[
 p(U)\ \longrightarrow\ \min\{J,\max\{-J,U-c\}\}.
\]
To see this, if \(p_*\) is a subsequential limit and
\(g_k(U-p_*)-g_k(c)>0\), then \(R_\epsilon\) tends to \(+\infty\)
and \(p_*=J\). A negative difference gives \(-J\).
A zero difference forces \(p_*=U-c\). These conditions select
the displayed limit.

The root field is \(A_d(U)-\sum_i p_i(U)\). In the limiting expression,
choose disjoint active intervals \([c_i-J_i,c_i+J_i]\).
On them the slope is \(A_d'(U)-1<0\), and on gaps it is
\(A_d'(U)>0\), since
\(A_d'(U)=d\cosh U/\sqrt{1+d^2\sinh^2U}\in(0,1)\).
On any fixed compact interval both slopes are
bounded away from zero in magnitude. Starting at its left end,
choose the \(N\) active intervals and \(N-1\) intervening gaps
so that the decreases and increases all have the same small
positive magnitude. The alternating endpoint heights, together
with the two tails, give \(2N+1\) crossings of a horizontal
level. The strict signs at finitely many test points persist
for a finite sufficiently small \(\epsilon>0\), hence at finite
couplings. A small shift of the root field to a regular value
preserves these crossings and makes all the roots simple by the
invertible eliminations above. This gives at least \(2N+1\)
roots in a slice of a positive-direction response that is a
single line.

\section{Realizing arbitrary genus in two-field response}
\label{sec:topology}

The construction starts with two circles and adds one fold at a time.
At the limiting scale the new equation doubles a sublevel surface
along its boundary. The limiting equations will be implemented by
stationary equations of a symmetric potential, so each step is
compatible with the Bethe gradient structure.

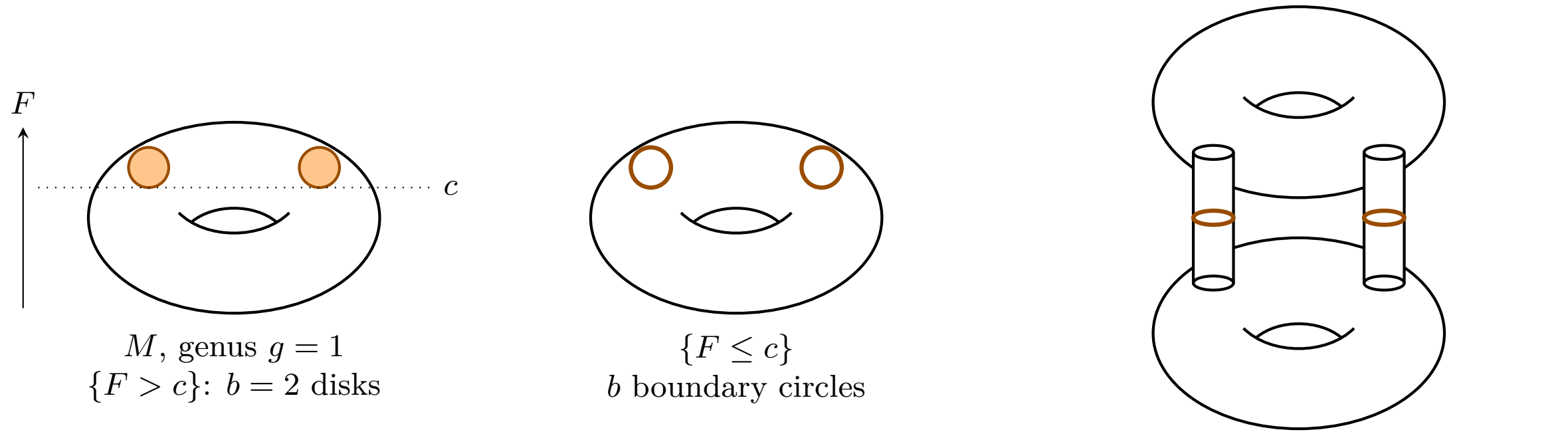
\begin{figure}[tbp]
  \centering
\providecommand{\schematictorus}[2]{%
  \draw[thick] (#1,#2) ellipse (1.45 and 0.95);
  \draw[thick] ({#1-0.55},{#2+0.05}) .. controls ({#1-0.28},{#2-0.22}) and ({#1+0.28},{#2-0.22}) .. ({#1+0.55},{#2+0.05});
  \draw[thick] ({#1-0.42},{#2-0.04}) .. controls ({#1-0.2},{#2+0.14}) and ({#1+0.2},{#2+0.14}) .. ({#1+0.42},{#2-0.04});
}
\begin{tikzpicture}[x=1cm,y=1cm,font=\footnotesize,>=stealth]
  \begin{scope}
    \schematictorus{0}{0}
    \fill[orange!45] (-0.85,0.5) circle (0.2);
    \fill[orange!45] (0.85,0.5) circle (0.2);
    \draw[thick,orange!60!black] (-0.85,0.5) circle (0.2);
    \draw[thick,orange!60!black] (0.85,0.5) circle (0.2);
    \draw[->] (-2.1,-0.9) -- (-2.1,0.9) node[above] {\(F\)};
    \draw[dotted] (-1.95,0.3) -- (1.95,0.3) node[right] {\(c\)};
    \node[align=center] at (0,-1.5) {\(M\), genus \(g=1\)\\ \(\{F>c\}\): \(b=2\) disks};
  \end{scope}
  \begin{scope}[xshift=5cm]
    \schematictorus{0}{0}
    \fill[white] (-0.85,0.5) circle (0.2);
    \fill[white] (0.85,0.5) circle (0.2);
    \draw[very thick,orange!60!black] (-0.85,0.5) circle (0.2);
    \draw[very thick,orange!60!black] (0.85,0.5) circle (0.2);
    \node[align=center] at (0,-1.5) {\(\{F\le c\}\)\\ \(b\) boundary circles};
  \end{scope}
  \begin{scope}[xshift=10.6cm]
    \schematictorus{0}{1.15}
    \schematictorus{0}{-1.15}
    \foreach \x in {-0.85,0.85} {
      \fill[white] (\x,0.65) circle (0.2);
      \fill[white] (\x,-0.65) circle (0.2);
      \fill[white] ({\x-0.2},-0.65) rectangle ({\x+0.2},0.65);
      \draw[thick] ({\x-0.2},-0.65) -- ({\x-0.2},0.65);
      \draw[thick] ({\x+0.2},-0.65) -- ({\x+0.2},0.65);
      \draw[thick] (\x,0.65) ellipse (0.2 and 0.07);
      \draw[thick] (\x,-0.65) ellipse (0.2 and 0.07);
      \draw[very thick,orange!60!black] (\x,0) ellipse (0.2 and 0.07);
    }
    \node[align=center] at (0,-2.65) {\(\widetilde M_c\): two copies glued along \(b\) circles\\ genus \(2g+b-1=3\)};
  \end{scope}
\end{tikzpicture}
  \caption{Doubling a sublevel surface. Left: a surface \(M\) of
  genus \(g=1\) with a height \(F\); the excursion set \(\{F>c\}\) above
  the level \(c\) consists of \(b=2\) disks. Middle: the sublevel set
  \(\{F\leq c\}\), a surface with \(b\) boundary circles. Right: the
  double \(\widetilde M_c\) of~\eqref{eq:double}, two copies of the
  sublevel set glued along the \(b\) circles. Its Euler characteristic
  is \(2(2-2g-b)\) and its genus is \(2g+b-1\), here \(3\).
  The sign of the new coordinate \(Z\) distinguishes the two copies.}
  \label{fig:doubling}
\end{figure}

\paragraph{Overview of the construction.}
The proof of Theorem~\ref{thm:C} occupies this section and the next,
and proceeds in five steps.
\begin{enumerate}
\item Two folds with one control produce a regular circle in a cubic
  stationary system of the form~\eqref{eq:circuit}. Two such circles
  with separate controls give a torus \(M_0\), whose height \(F_0\) is
  the sum of the two controls (Section~\ref{sec:two-folds}).
\item Adding one fold variable with the equation \(Z^2=c-F\) replaces a
  surface \(M\) by the double of its sublevel set \(\{F\leq c\}\). If the
  excursion set \(\{F>c\}\) consists of \(b\) disks, the genus becomes
  \(2g+b-1\) (Figure~\ref{fig:doubling}, Section~\ref{sec:doubling}).
\item Iterating the doubling along a chain of \(k\) folds, each coupled
  only to its predecessor, realizes every genus from \(0\) to
  \(3\cdot2^k-1\) with one set of mixed coefficients. Only the linear
  coefficients depend on the genus (Section~\ref{sec:chain}).
\item The cubic system is the scaling limit of BP models in which each
  fold variable is a \(K_{2,3}\) block at its fold and each mixed term
  is a weak edge (Section~\ref{sec:cubic-to-bp},
  Appendix~\ref{app:wiring}). This already gives the surfaces, with
  varying edge strengths.
\item Section~\ref{sec:uniform} replaces the weak edges by two-terminal
  blocks with the single coupling \(J_0\), whose effective mixed
  coefficient is tuned through zero by a negative internal mode, and
  proves the matching size lower bound.
\end{enumerate}

\subsection{A circle from two folds}
\label{sec:two-folds}

Use \(p\) fold coordinates \(X\) and two control coordinates \(Y\).
The potentials used in the construction have the form
\begin{equation}
 V(X,Y)=\sum_i\frac{\sigma_i X_i^3}{3}
  -\sum_{i<j}\lambda_{ij}X_iX_j-\sum_{i,\alpha}a_{i\alpha}X_iY_\alpha
  -\sum_i t_i X_i,\qquad
 \sigma_i\in\{-1,1\},\quad \lambda_{ij},a_{i\alpha}\geq0 .
 \label{eq:circuit}
\end{equation}
Only the equations \(\partial_X V=0\) are imposed.
The signs \(\sigma_i\) select the curvatures of the folds.
The nonnegative coefficients of the mixed terms will correspond
to positive couplings in the first realization.

Consider two folds with signs \(-,+\), one control, and
\(\lambda=L/\sqrt2\), \(t=L^2/8-1/2\). Choose control coefficients
\(1,1\) and linear coefficients \(t,-t\).
After multiplying the first equation by \(-1\) the system is
\[
 X_0^2+\lambda X_1+Y+t=0,\qquad
 X_1^2-\lambda X_0-Y+t=0 .
\]
Put \(x=X_0-\lambda/2\), \(z=X_1+\lambda/2\).
Adding the equations gives \(x^2+z^2=1\), and one equation solves
for \(Y\). With \(a=(x+z)/\sqrt2\), \(b=(z-x)/\sqrt2\), the control
coordinate is
\[
 Y=f_L(a,b)=a(b-L),\qquad a^2+b^2=1.
\]
The circle is regular, since the sum has nonzero differential on
the circle and the other equation has nonzero \(Y\) derivative.

Take two copies with \(L_1=1/8\), \(L_2=1/16\), with separate
controls. The product of their circles is a torus \(M_0\) and
the height \(F_0=Y_1+Y_2\) is \(f_{L_1}+f_{L_2}\).
On a circle, using \(a=\cos\theta\), \(b=\sin\theta\), one has
\(f_L'=1-2b^2+Lb\). For these \(L\) its two roots are strictly
between \(-1\) and \(1\), and each has two preimages. Thus
each height has two nondegenerate maxima and two nondegenerate
minima, the minima being the negatives of the maxima.
Let \(M_1<M_2<M_3<M_4\) be the four sums of two maximum
values. The radical formulas in Appendix~\ref{app:wiring} give
\[
 \tfrac45<M_1<\tfrac9{10}<M_2<1<M_3<\tfrac{11}{10}
 <M_4<\tfrac54 .
\]
The mixed sums of a maximum and a minimum have absolute
value less than \(1/4\), so the extreme levels used below are
separated from all saddles.

\subsection{Doubling a sublevel surface}
\label{sec:doubling}

Let \(M\) be a regular compact connected component of a system
of the form~\eqref{eq:circuit}. Let \(F\) be an affine function
of \((X,Y)\) with nonnegative linear coefficients, whose
restriction to \(M\) is Morse. For a regular value \(c\), set
\begin{equation}
 \widetilde M_c=\{(x,Z):x\in M,\ Z^2=c-F(x)\}.
 \label{eq:double}
\end{equation}
The two signs of \(Z\) give the two copies of \(\{F\leq c\}\),
joined smoothly at \(Z=0\). At those points \(d(F|_M)\ne0\).
If \(M\) has genus \(g\) and \(\{F>c\}\) consists of \(b\geq1\)
disjoint open disks, then the double is connected and
\[
 \chi(\widetilde M_c)=2(2-2g-b),\qquad
 \operatorname{genus}(\widetilde M_c)=2g+b-1 .
\]
When \(c>\max_M F\) there are two disjoint copies of \(M\).
Either copy can be retained as a component.

Suppose the two highest critical values are distinct and
strictly above all other critical values, each attained at one
local maximum. Make the analogous assumption on the two lowest
values and minima. There are cuts for \(b=1\) and \(b=2\),
since the upper excursion sets immediately below the first
and second maximum levels consist of one and two disks,
respectively. Together with the copy operation this gives
\(g,2g,2g+1\). The new height \(Z\) can be used again.
At \(Z=0\) its differential is nonzero on the double.
Elsewhere its critical points come from those of \(F\), and at
a critical point
\(\operatorname{Hess} Z=-(2Z)^{-1}\operatorname{Hess} F\)
on the corresponding sheet. In a connected double the two
lowest values of \(F\) produce the two highest values of \(Z\)
on the positive sheet and the two lowest on the negative sheet.
On a positive copy the highest and lowest pairs exchange roles.
Thus the conditions at the extreme levels persist.
All these conditions are open under sufficiently small \(C^2\)
perturbation of the surface and the height.

To implement~\eqref{eq:double} while keeping symmetric mixed terms,
add a new fold variable \(U\) to \(V\) by replacing it with
\[
 V_r=V-\frac{U^3}{3}-r^2U(F-c).
\]
The new potential has the form~\eqref{eq:circuit}. Write \(U=rZ\)
and \(a=\partial_X F\). Its conditional equations, after a nonzero
row rescaling, are
\[
 \partial_X V-r^3 Za=0,\qquad Z^2+F-c=0 .
\]
As \(r\to0\) this converges in \(C^q\) on compact sets for any
fixed \(q\) to the system defining~\eqref{eq:double}.
At \(Z\ne0\) the new \(Z\) derivative gives full row rank.
At \(Z=0\), full row rank follows from regularity of \(c\) on
\(M\). Lemma~\ref{lem:persistence} gives an isolated component of the
same genus for sufficiently small positive \(r\), and its
height \(U/r\) inherits the conditions at the extreme levels.
The perturbation can be taken small for any finite list of cuts
and components at once.

\subsection{A fixed chain for a range of genera}
\label{sec:chain}

Start from \(M_0\) and add the first fold \(U_1\) with \(r_1=1\)
and \(F=F_0\). Since \(F_0\) is a sum of control variables, this
operation leaves the first four equations exactly unchanged.
At the cuts \(c_1=5/4,11/10,1,9/10,4/5\), the upper excursion
set removes \(0,1,2,3,4\) disks, respectively. Choosing the positive
copy when \(b=0\) realizes genera \(1,2,3,4,5\). In each connected
double the two lowest critical values of \(F_0\) remain, giving
the two separated extremes of each sign for the height \(U_1\).
The cut \(c_1=-11/10\) lies strictly between the two lowest
critical values. Its sublevel set is a disk, whose double is
a sphere. For that setting the height is Morse with two critical
points.

Thereafter each new fold connects only to its predecessor:
\[
 V_j=V_{j-1}-\frac{U_j^3}{3}
      -r_j^2 U_j\left(\frac{U_{j-1}}{r_{j-1}}-c_j\right).
\]
Once the finite set of cuts for all previous settings has been
chosen, one sufficiently small \(r_j>0\) works for all of them.
For every previous positive genus \(g\), choose cuts for the
copy, \(2g\), and \(2g+1\). For the sphere always take the positive
copy. Thus from all genera \(0,\ldots,m\) we obtain
\(0,\ldots,2m+1\). With \(m=5\) at the first step, the chain of
\(k\) added folds realizes every genus from zero to \(3\cdot2^k-1\).
All mixed coefficients are common across the resulting finite
list of settings. Only the linear coefficients of the added folds
depend on the cuts. More precisely, for each fixed \(k\) the
induction produces a finite set of linear coefficient vectors
\(\mathcal T_k\) and one common collection of mixed coefficients.
Each selected component has its own compact isolating tube.
At each induction step take the minimum of the finitely many
positive persistence tolerances. This is the simultaneous choice
of \(r_j\) used above; it may depend on \(k\).

\subsection{From the cubic equations to BP}
\label{sec:cubic-to-bp}

The cubic equations can be obtained from finite, strictly
ferromagnetic BP models by using ordinary folds as the individual
blocks. At a fold, the reduced potential in the kernel direction
starts with a cubic of nonzero coefficient. A weak positive edge
between two blocks supplies a negative mixed quadratic term
in their kernel directions. An edge to a controlled vertex
supplies the mixed term with its magnetization. The noncontrolled
fields are centered at the exact field map of the coupled model
at the reference magnetizations. Field offsets then give the
linear terms of~\eqref{eq:circuit}.

Appendix~\ref{app:wiring} gives the ordinary folds and the scaling
calculation. In the three-port implementation, each \(X_i\) uses
a \(K_{2,3}\) block with three available attachment vertices, and
each control is a single vertex. Our chain needs at most three
attachments at any block or control, so this realizes the chain
on a connected simple graph of maximum degree three. At a fixed
chain length the weak couplings can be chosen simultaneously
for the whole finite list of linear coefficients. The full
noncontrolled equations have full row rank on each persisted
component, by the invertible elimination and regularity of
\(\partial_X V=0\) there. The controlled fields are then free,
with the prescribed noncontrolled fields held fixed. This gives
the desired surfaces with varying edge strengths. The next
section implements the same cubic potential with a common edge
strength throughout the graph.

\section{Uniform couplings, bounded fields, and optimal size}
\label{sec:uniform}

Throughout the construction in this section each physical edge has
the coupling \(J_0\) of Theorem~\ref{thm:C}. Write \(t=\tanh J_0\).
The weak mixed coefficients of the cubic potential must now come
from stationary elimination. We construct a bounded-field
two-terminal block whose effective mixed Hessian entry can be
adjusted through zero. The elimination uses a branch with one
negative internal mode.

\subsection{Stationary elimination at zero terminal magnetization}

For an attached block, fix the exterior terminal magnetizations at
zero and specify reference interior magnetizations. Set the interior
fields to the corresponding values of the full field map, and
eliminate the interior stationary equations near this point when
their Hessian block is invertible. Let \(\Psi(u,v)\) be the resulting
contribution to the exterior potential, excluding the two exterior
one-vertex terms. Its quadratic part is obtained by a Schur complement.
The normalized Hessian \(A\) in~\eqref{eq:hessian} gives the same
exterior coefficients after elimination, since the exterior reference
magnetizations are zero.

We use two elementary blocks, whose calculations are in
Appendix~\ref{app:uniform}.
The first is a simple bipartite transmission block with \(18\)
interior vertices and opposite-color exterior terminals, each using
one edge. The interior Hessian is positive definite. By fixing
suitable interior magnetizations close to saturation, one can make
the norm of the exterior Hessian contribution arbitrarily small.
The exterior mixed term is strictly negative at any such finite
setting. An interior symmetry followed by spin reversal fixes the
interior fields and reverses both exterior magnetizations, so
\(\Psi(-u,-v)=\Psi(u,v)\).

The second block is \(K_{2,3}\) with three zero-magnetization
degree-two ports. When the other two magnetizations are \((x,-x)\)
or \((x,x)\), all six edge correlations are the same \(b(x)>0\),
and \(b(x)^2\) decreases from \(t^2>1/2\) to zero for \(x\in(0,1)\).
There is a simple crossing of the lowest Hessian eigenvalue at
\(x=x_f\), where \(b(x_f)^2=1/2\), with a positive kernel vector
and four positive complementary eigenvalues.
For the same-sign choice \((x_f,x_f)\), the cubic derivative
in the kernel direction is strictly positive. The spin-reversed
choice gives a strictly negative derivative. Both properties
persist at the shifted crossing after small diagonal port
corrections.

\subsection{Cancellation by a negative internal mode}

\begin{lemma}[Tunable two-terminal block]
\label{lem:cancellation}
For every fixed positive tolerance on the two exterior diagonal
contributions, a simple bipartite block with \(120\) interior
vertices realizes every sufficiently small mixed Hessian entry
of either sign, including zero. The exterior terminals have
opposite colors and use one edge each, and all interior vertices
have degree three after attachment. The interior reference fields
stay in a common finite box. The stationary elimination is
regular and its contribution is even under simultaneous reversal
of the two exterior magnetizations.
More precisely, after fixing the tolerance there are \(\eta>0\)
and a compact interval of reference settings on which all internal
magnetizations stay strictly between \(-1\) and \(1\), the internal
Hessian is invertible, and the mixed entry takes every value in
\((-\eta,\eta)\). The eliminated potentials have uniformly bounded
state derivatives of each fixed finite order on a common
neighborhood of the exterior origin.
\end{lemma}

\begin{proof}
Use two oppositely colored copies of \(K_{2,3}\), joining two
ports of each to their counterparts through two of the stable
transmission blocks. The left internal pair has magnetizations
\((x,-x)\); on the right use \((x_R,-x_R)\) with \(x_R>x_f\)
fixed on the positive definite side. The remaining port of
each copy is joined, through another stable transmission block,
to a new junction vertex \(j_1\) or \(j_2\). The two junctions
have opposite colors and share a direct physical edge. All
ports and junctions have zero reference magnetization.

After eliminating the four stable transmission blocks, let
\(C(x)\) be the normalized Hessian on the ten vertices of the
two copies of \(K_{2,3}\), \(B\) the two junction-to-core rows,
and \(A_0\) the junction block. The rows of \(B\) have one
strictly negative entry each. The direct edge gives
\((A_0)_{12}=-t/(1-t^2)\), and \(A_0\) is positive definite
when the transmission contributions are sufficiently small.
At zero transmission contribution the left core block crosses
simply at \(x_f\) and the right is positive definite. Fix a
small closed interval about \(x_f\). For sufficiently small
transmission contributions the crossing persists at some \(x_*\)
inside this interval. Below \(x_*\) the core has one negative
eigenvalue \(\lambda(x)\) and a uniform positive gap for the
other eigenvalues. The connected core is a symmetric matrix with
nonpositive off-diagonal entries, so its lowest unit eigenvector
\(q(x)\) is strictly positive up to the crossing.

Below the crossing the junction Schur complement is
\begin{equation}
 D(x)=A_0-BC(x)^{-1}B^\top
 =A_0-BC_+(x)^{-1}B^\top
   +\frac{(Bq(x))(Bq(x))^\top}{|\lambda(x)|},
 \label{eq:resonance}
\end{equation}
where \(C_+^{-1}\) is zero on the negative eigenspace and inverse
on the positive eigenspace. The first two terms together are
uniformly positive definite by the gap and the smallness of
\(B\). At the left endpoint of the interval \(D_{12}<0\),
since the core inverse stays bounded there and the direct
edge dominates. As \(x\uparrow x_*\), \(D_{12}\to+\infty\):
both entries of \(Bq\) are strictly negative at the crossing.
There is a sign-changing zero \(x_s<x_*\) of this analytic
entry. Choose a closed interval about \(x_s\) contained strictly
in the negative-eigenvalue side of the crossing. On a sufficiently
small such interval all elimination blocks are invertible and
their inverses are uniformly bounded. In particular, the chosen
interval does not approach the pole at \(x_*\).

Use two more stable transmission blocks to attach \(j_1,j_2\)
to the actual exterior terminals, with mixed coefficients
\(-\mu_1,-\mu_2<0\). Their junction contributions change only
the diagonal of \(D\). Eliminating the junctions gives the mixed
entry
\begin{equation}
 -\mu_1\mu_2(D^{-1})_{12}
   =\frac{\mu_1\mu_2D_{12}}{\det D}.
 \label{eq:effective}
\end{equation}
It changes sign through zero at \(x_s\), and admits a continuous
local inverse branch through the crossing, since the zero has
odd order. The exterior diagonal contributions can be made
arbitrarily small by choosing the last two transmissions small:
\(D^{-1}\) is uniformly bounded on the interval and the
exterior-to-junction rows are small. The inertia formula for
Schur complements gives exactly one negative direction in the
entire eliminated interior Hessian.

The vertex count is \(10+2+6\cdot18=120\).
Each of the ten core vertices and the two junctions has degree
three. The two internal pairs can be exchanged and the two
halves of each stable transmission swapped, in an automorphism
fixing the exterior terminals and negating the reference
magnetizations. The automorphism composed with spin reversal
preserves the fields at that reference. Uniqueness of the chosen
local elimination branch gives the required evenness. All
interior magnetizations used are in a fixed compact subset
of \((-1,1)\), including as the mixed entry tends to zero.
The parameter-dependent analytic implicit function theorem on
the compact interval gives a common exterior neighborhood and
uniform state-derivative bounds. The odd-order zero and its
continuous inverse branch give the interval \((-\eta,\eta)\)
in the statement. The field bound follows from~\eqref{eq:proper}.
\end{proof}

\subsection{Realization of the cubic potential at one coupling}

Fix one sufficiently small diagonal tolerance and one corresponding
family from Lemma~\ref{lem:cancellation}, before choosing \(k\).
Each terminal has at most three attachments. Thus a fixed tolerance
can keep every middle-vertex diagonal above \(1/2\) and every active
block within the neighborhood where its simple crossing and nonzero
cubic coefficient persist. Restrict the mixed-entry interval further
if necessary. All copies below use this same compact family.

Take the \(p=k+4\) blocks and \(p+3\) logical edges of
Section~\ref{sec:topology}. Put the three ports of each \(K_{2,3}\)
and the two control vertices in the same color class. Replace
each logical edge \(e\) by a path through a new vertex of the
opposite color, using one cancellation block on each leg.
These middle vertices have one remaining attachment each.
Use \(p\) of those attachments to connect to the \(p\) unused
ports, one each, through cancellation blocks at mixed
coefficient zero. Use the remaining three to attach to one
new vertex in the ports' color class, again through blocks
at mixed coefficient zero. This is a connected simple
bipartite cubic graph. There are \(6p+6\) vertices outside
the cancellation blocks and \(3p+9\) such blocks. Hence the
number of vertices is
\begin{equation}
 6p+6+120(3p+9)=366k+2550 .
 \label{eq:vertices}
\end{equation}

All endpoints of the cancellation blocks are centered at
magnetization zero. After the blocks are eliminated, the
middle vertex of a logical edge \(e\) has Hessian diagonal
\(d_e>1/2\) by the small diagonal tolerances. Its mixed entries
with the two endpoints are \(\xi_{e,1},\xi_{e,2}\).
At second order the extra degree-completing links transmit
nothing, so the middle vertices and the new degree-three
vertex have an invertible diagonal block. Eliminating them
creates, for each logical edge, the mixed term
\begin{equation}
 -\frac{\xi_{e,1}\xi_{e,2}}{d_e}
 \label{eq:logical}
\end{equation}
between its original endpoints. The ancillary potential after
these eliminations is even in all remaining port and control
magnetizations jointly, since the eliminated vertices are
centered at zero and their centered potential is even.
Thus its third derivatives vanish at the reference point.
Its diagonal corrections at the ports are small.

For each active \(K_{2,3}\), retune its same-sign internal pair
to the positive or negative crossing after adding the actual
diagonal port corrections. The full Hessian block of this
individual block has one zero eigenvalue and four positive
ones. Choose its positive kernel vector \(v_i\) scaled so
that the cubic derivative in that direction is \(2\sigma_i\)
with the desired sign. This is possible by the calculation
in Appendix~\ref{app:uniform}. Denote by \(v_i^0\) its value
when every cancellation block has mixed entry \(\xi=0\), and
by \(d_e^0\) the corresponding middle-vertex diagonal. At this
setting the stable transmission blocks are held fixed at finite
magnetization, and the cancellation parameter is \(x_s\).
Its small diagonal corrections are retained in \(v_i^0,d_e^0\).
Let \(w_e>0\) be the product of the
two relevant port components of \(v_i^0\), with value \(1\)
on the control side of a control edge. Let \(\lambda_e>0\)
denote the target coefficient, either \(\lambda_{ij}\) or
\(a_{i\alpha}\) in~\eqref{eq:circuit}.

Choose a small magnetization scale \(\epsilon>0\). Use
Lemma~\ref{lem:cancellation} to set
\[
 \xi_{e,1}=\xi_{e,2}
   =\sqrt{\epsilon\,\lambda_e d_e^0/w_e}.
\]
Then the coefficient in~\eqref{eq:logical} projected to the
kernel and control directions is
\(-\epsilon\lambda_e+o(\epsilon)\), since the diagonal
corrections, kernel vectors, and middle-vertex diagonals
converge to their values at the finite cancellation setting.
For each fixed \(k\) there are finitely many target coefficients,
so one \(\epsilon\) puts all these mixed entries in the common
interval \((-\eta,\eta)\). The unused port links remain at zero.

Let \(m^*\) collect the chosen reference magnetizations.
Use the exact field \(H_{J_0,W}(m^*)\) and add on block \(i\)
the shift \(\epsilon^2 t_i a^{(i)}\), where
\(v_i^\top a^{(i)}=1\). We can choose \(a^{(i)}\) supported on
the two internal vertices of the block, with a uniform norm
bound: the positive kernel coordinates there remain bounded
away from zero on the fixed compact fold neighborhoods.
Put the control
magnetizations at \(\epsilon Y_\alpha\) and the active block
magnetizations at \(m_i^*+\epsilon X_i v_i+z_i\).
The complementary Hessians have invertible limits.
The scaled equations for \(z_i=\epsilon^2 Z_i\), divided by
\(\epsilon^2\), have these invertible blocks as their limiting
\(Z\) derivative. The forcing consists of the \(O(\epsilon)\)
mixed entries acting on \(O(\epsilon)\) displacements, the
quadratic gradient terms, and the prescribed field shifts.
The complementary directions can therefore be eliminated
with \(z_i=O(\epsilon^2)\), including in any fixed number of
state derivatives on compact coordinate sets.

Subtract the constant and terms involving only \(Y\) from
the reduced potential, and divide by \(\epsilon^3\).
The individual quadratic terms in \(X\) vanish by the fold
adjustment. The cubic derivatives give \(\sigma_i X_i^3/3\),
the mixed Hessian entries give the desired mixed terms, and
the field shifts give \(-t_iX_i\). The complementary
elimination affects the energy at order \(O(\epsilon^4)\).
Fourth and higher terms are also \(O(\epsilon^4)\) on
compact scaled sets. The eliminated blocks have uniformly
bounded state derivatives near their reference settings,
and their even contribution has zero cubic part. These
estimates yield convergence to~\eqref{eq:circuit} in \(C^q\)
for every fixed finite \(q\). The coefficients of the block
settings need only converge with \(\epsilon\), as provided by
the continuous inverse branches.

The regular compact components of the limiting equations
persist at sufficiently small positive \(\epsilon\), by
Lemma~\ref{lem:persistence}. The invertible elimination blocks
make the differential of the original noncontrolled equations
equivalent, by invertible row and coordinate operations, to
those blocks together with the differential of the reduced
conditional equations. The latter has full row rank on each
persisted surface. Reconstruction is an embedding into a local
coordinate neighborhood of the full positive family. The resulting
surface is compact and locally exhausts the zero set, and hence
is a connected component of the full response.

For a fixed \(k\), apply this argument to the finite set
\(\mathcal T_k\) and its compact isolating tubes from
Section~\ref{sec:topology}. Taking the minimum of their persistence
tolerances gives one \(\epsilon>0\) for all settings. The mixed
coefficients, reference magnetizations, graph, and controls are
common to this list. Only the shifts in \(t_i\) change with the
desired genus.

There is a universal \(m_{\max}<1\) bounding all absolute
reference magnetizations. The stable transmission settings
and compact interval about \(x_s\) were fixed before \(k\).
The at most three small port corrections keep each active
pair in fixed compact neighborhoods of the two fold values.
The remaining reference magnetizations are zero. Consequently
the same \(m_{\max}\) applies to every number of copies of these
blocks. Thus~\eqref{eq:proper} bounds
the reference fields by \(\atanh m_{\max}+3J_0\).
For each finite list, further decrease \(\epsilon\) until
\(\epsilon^2\max_{t\in\mathcal T_k,i}
\bigl(|t_i|\,\|a^{(i)}\|_\infty\bigr)\leq1\).
This is compatible with the preceding
persistence and mixed-entry conditions. Therefore
\(H_0=\atanh m_{\max}+3J_0+1\) works for all \(k\).
Together with~\eqref{eq:vertices}, this proves the upper
bound and the stated finite list of genera in
Theorem~\ref{thm:C}.

\subsection{The constructed surfaces consist of Bethe saddles}

\begin{lemma}[Negative index on the constructed surfaces]
\label{lem:negative-index}
Let \(k\geq1\) and let \(\Sigma\) be one of the regular compact
components produced above on the graph with \(366k+2550\) vertices,
for any genus \(0\leq g\leq3\cdot2^k-1\). At every point of
\(\Sigma\), the Hessian \(K=D_mH_{J_0}\) of the reduced Bethe
potential has at least \(3k+21\) negative eigenvalues. In particular
no point of \(\Sigma\) is a local minimum of the Bethe free energy.
\end{lemma}

\begin{proof}
The construction uses \(3p+9=3k+21\) cancellation blocks. Let \(I\)
be the union of their interior vertex sets. Interior vertices of
different blocks are not adjacent, since each block meets the rest of
the graph only through its two terminals. Hence the principal
submatrix \(K_{II}\) is block diagonal with one block per
cancellation gadget, and each block is the interior Hessian of that
gadget at the current magnetizations. On the compact family of
Lemma~\ref{lem:cancellation} the interior Hessian is invertible with
exactly one negative eigenvalue, by the inertia count in its proof.
Points of \(\Sigma\) have all magnetizations within \(O(\epsilon)\)
of their reference settings, uniformly over the finite list
\(\mathcal T_k\), by the scaled ansatz and the elimination
\(z=O(\epsilon^2)\). Since \(K\) is continuous in \(m\) and the
interior Hessians of the compact family have uniformly bounded
inverses, a further decrease of \(\epsilon\), compatible with the
previous conditions, keeps exactly one negative eigenvalue in each
block on \(\Sigma\). Thus \(K_{II}\) has \(3k+21\) negative
eigenvalues, and by Cauchy interlacing for principal submatrices
\(K\) has at least as many. The normalized matrix
in~\eqref{eq:hessian} is congruent to \(K\) by a positive diagonal
matrix and has the same count. The edge directions eliminated from
the Bethe potential are positive definite, so the negative index of
\(K\) is that of the full Bethe Hessian at the fixed point, as noted in
Section~\ref{sec:field-map}.
\end{proof}

\begin{remark}
Lemma~\ref{lem:negative-index} describes the mechanism of the
construction. The topology is produced by cancelling positive
transmissions against the response of internal negative modes, and
every point of the resulting surface is a saddle of the Bethe free
energy. Heskes \cite{heskes} proved that stable fixed points of loopy
BP, in the sense of that paper, are local minima of the Bethe free
energy. Under that correspondence, none of the points on these
surfaces is a stable fixed point of the iteration. We state the
result for the negative index rather than the Morse index because
\(K\) is singular at the critical points of the field projection.
The bound is specific to this construction. We do not claim it for
every model that realizes a given genus.
\end{remark}

\subsection{An algebraic upper bound on Betti numbers}

We prove the matching lower bound on vertex count by
bounding the number of Morse critical points on a response
component. Let \(D\) be the two controls. Fix \(a\) and \(b_W\)
and use coordinates \(\zeta=(y,b_D)\) for the positive response.
Let \(r=2|E|\), so there are \(r\) BP polynomials in \(r+2\)
variables. A polynomial belonging to an arc with tail \(i\)
has total degree at most
\(\delta_\alpha=\deg(i)+\mathbf 1_{i\in D}\).
On a regular compact component the Jacobian of these
polynomials has full row rank \(r\): the positive family
has the local analytic presentation of
Section~\ref{sec:field-map}, and fixing \(b_W\)
is transverse by regularity. Deleting these fixed coordinate
columns and the equations fixing them gives the asserted rank
in \(\zeta\) coordinates.

A generic real linear height \(c^\top\zeta\) is Morse on the
component, by Sard's theorem applied to the normal bundle.
At a critical point there is a unique multiplier \(\lambda\)
with \(f=0\) and \(c=Df^\top\lambda\). The Jacobian of this
system in \((\zeta,\lambda)\) is invertible: its kernel first
requires a tangent displacement, and then the nondegenerate
tangent Hessian forces that displacement to vanish. Full row
rank of \(Df\) then forces the multiplier displacement to
vanish as well. Thus every Morse point supplies an isolated
simple complex solution of the critical equations.

Perturbing the polynomials and the linear height to general
coefficients of the same degree bounds preserves these finitely
many simple roots,
even if the original complex critical equations have other
components. For general equations and a linear objective
the number of complex critical points is
\((\prod_\alpha\delta_\alpha)h_2(\delta_\alpha-1:\alpha\in\mathcal A)\),
where \(h_2\) is the complete homogeneous polynomial of degree
two. Here the ambient dimension minus the number of equations
is two. The perturbation is in polynomial coefficient space.
\cite[Theorem~2.2]{polar}

For an \(n\)-vertex cubic graph there are \(3n\) arc polynomials,
six of degree bound four and the others of degree bound
three. Using \(h_2(x)=((\sum x)^2+\sum x^2)/2\), Morse
inequalities for a connected orientable closed surface give
\[
 2+2g\ \leq\ 4^6\,3^{3n-6}\,(18n^2+42n+33).
\]
This forces \(n=\Omega(\log(g+1))\) for \(g\geq1\).
Taking \(k\) of order \(\log(g+1)\) in the construction
gives the matching upper bound and completes
Theorem~\ref{thm:C}. The quantity \(n_{\min}(g)\) counts vertices
only. The construction does not control, uniformly in \(g\), the
precision with which the fixed fields must be specified, the scale
\(\epsilon\), or the condition number of the response, and
Theorem~\ref{thm:C} makes no claim about them.

\section{Certified bifurcation diagrams and topology from root counts}
\label{sec:diagrams}

We first compute the entire one-field response of a specific
nine-vertex model. The product of two copies gives the complete
two-field diagram in Figure~\ref{fig:discriminant}. All counts in
this section are counts of distinct positive roots.

\subsection{The nine-vertex model}

Use two diamonds on vertices \(0,1,2,3\) and \(4,5,6,7\),
with endpoints \(0,1\) and \(4,5\), respectively.
The first has edges \(01,02,12,03,13\), and the second is its
translate by four. Give the direct edge of each diamond
coupling \(J_*=\tfrac14\log 10\) and the remaining edges
\(\tanh J_{\rm path}=\sqrt{\tanh J_*}\).
Connect a control vertex \(8\) to the four endpoints, with
coupling \(J_c=\tfrac12\log(1001/1000)\) on each of the four
new edges. The fixed fields at the first two endpoints
are \(H=\tfrac14\log(102400/100001)\), those at the second
two endpoints are \(-H\), and those at the four path
vertices are zero. The control field \(h=h_8\) is free.

\begin{proposition}[Complete one-field diagram]
\label{prop:one-field}
The positive response has three line components and one circle
\(C_1\), all regular. All critical points of its projection to
\(h\) are ordinary folds. Their six critical values are
\(\pm a,\pm b,\pm c\), with \(0<a<b<c\) and
\[
 \begin{aligned}
 0.004810&<\tanh a<0.004835,\\
 0.004836&<\tanh b<0.004860,\\
 0.005766&<\tanh c<0.005771 .
 \end{aligned}
\]
Table~\ref{tab:one-field} gives the root counts and Hessian
inertias on every field cell.
\end{proposition}

\begin{table}[htbp]
\centering
\small
\begin{tabular}{@{}lcccc@{}}
\toprule
Field cell & On \(C_1\) & All roots & \(\#C_1\) & Total \\
\midrule
\(\abs h<a\)     & \(1+2s+s^2\) & \(4+4s+s^2\) & 4 & 9 \\
\(\abs h=a\)     & \(s+s^2+z\) & \(3+3s+s^2+z\) & 3 & 8 \\
\(a<\abs h<b\)   & \(s+s^2\) & \(3+3s+s^2\) & 2 & 7 \\
\(\abs h=b\)     & \(sz\) & \(3+2s+sz\) & 1 & 6 \\
\(b<\abs h<c\)   & \(0\) & \(3+2s\) & 0 & 5 \\
\(\abs h=c\)     & \(0\) & \(2+s+z\) & 0 & 4 \\
\(\abs h>c\)     & \(0\) & \(2+s\) & 0 & 3 \\
\bottomrule
\end{tabular}
\caption{The nine-vertex model. Each monomial \(s^i z^j\)
counts one root with \(i\) negative and \(j\) zero Hessian eigenvalues.
Evaluation at \(s=z=1\) gives the distinct-root count.}
\label{tab:one-field}
\end{table}

\begin{proof}[Proof of Proposition~\ref{prop:one-field}]
The endpoint pair of each diamond has equal magnetizations at
every root. To see this, keep their mean fixed and vary their
difference. The difference of the fields vanishes when the
magnetizations agree, by symmetry. Its derivative in the
difference direction is strictly positive by~\eqref{eq:hessian},
since \(K_{ii}+K_{jj}-2K_{ij}>0\).
Eliminate the zero-field degree-two vertices as in
Appendix~\ref{app:wiring}. Let \(f\) be the common field of
three parallel factors in~\eqref{eq:core-field}, and put
\(\alpha=\tanh J_c=1/2001\), \(\eta=-H\).
Use coordinates \(u=-m_0=-m_1\), \(v=m_4=m_5\), and \(y=m_8\).
For either pair in the sign convention with fixed field \(\eta\), set
\[
 z_0(x)=\tanh(f(x)-\eta),\qquad \tau(x)=z_0(x)/\alpha,\qquad
 Y(x)=\frac{\tau(x)(1-\alpha^2)+\alpha x(1-\tau(x)^2)}
                   {1-z_0(x)^2}.
\]
The allowable interval condition is \(\abs{z_0(x)}<\alpha\).
In fact the field residual at the endpoint requires the incoming
message from the control to have half-log \(f(x)-\eta\).
This fixes the control cavity magnetization to \(\tau(x)\).
Taking the conditional means of the control spin given the
endpoint spin and averaging with mean \(x\) gives \(Y(x)\).
Conversely these equations reconstruct a positive edge table.

The endpoint cavity magnetization on the same edge is
\((x-z_0(x))/(1-xz_0(x))\).
Define
\[
 L(x)=2\atanh\left(\alpha\frac{x-z_0(x)}{1-xz_0(x)}\right).
\]
The exact full response and its field map are
\begin{equation}
 Y(u)+Y(v)=0,\qquad y=-Y(u)=Y(v),\qquad
 h=-\atanh Y(u)+L(u)-L(v).
 \label{eq:diagram-curve}
\end{equation}
All positive roots are recovered, including at the folds.
Interchanging \(u,v\) reverses \(h\).

The allowed set of \(x\) has two open intervals, denoted
\(R,F\). They are contained respectively in
\([-61/100,-59/100]\) and \([1/4,41/100]\).
This follows from the monotonicity of \(f\) in
\eqref{eq:f-derivative} and the certified endpoint evaluations in
Appendix~\ref{app:certificate}. The same certificate and
differentiation give
\begin{equation}
 \begin{gathered}
 Y'|_R>190,\qquad |Y''|_R<5000,\qquad
 200<Y''|_F<1100,\\
 0<L'<1/500,\qquad |L''|<1/50 .
 \end{gathered}
 \label{eq:diagram-bounds}
\end{equation}
On \(R\), \(Y\) increases from \(-1\) to \(1\).
On \(F\) it has value tending to \(1\) at both ends,
and has a unique minimum \(Y_*=Y(x_*)\) with
\(-1/190<Y_*<-1/250\). On the ends of
\([8/25,26/75]\subset F\) the value is above \(1/50\).
In \(F\times F\) the curve in~\eqref{eq:diagram-curve}
therefore bounds a strongly convex disk inside that square.
This is \(C_1\). In \(R\times R\) it is a graph over
\(y\in(-1,1)\), and \(F\times R\), \(R\times F\) are graphs
over the \(F\) coordinate. This gives three lines and one
circle. The curve differential never vanishes: both
\(Y'(u),Y'(v)\) would vanish only at \((x_*,x_*)\), which is
outside the zero level. The earlier eliminations were
regular, so the full conditional equations also have
full row rank.

Put \(p=Y'(u)\), \(q=Y'(v)\). In tangent direction
\((-q,p)\) the field derivative is
\[
 \frac{pq}{1-Y(u)^2}-L'(u)q-L'(v)p .
\]
If both \(|p|,|q|>1/250\) it cannot vanish, since after division
by \(pq\) the last two terms have absolute sum less than \(1\).
Consider first \(|p|\leq 1/250\).
Then \(u\) is near \(x_*\) in \(F\), and integrating
\(d(p^2)=2Y''\,dY\) along the branch gives
\[
 0\le Y(u)-Y_*\le p^2/400 .
\]
Thus \(Y(v)=-Y(u)>1/250-1/(400\cdot250^2)\).
In \(F\) this gives \(q^2\geq400(Y(v)-Y_*)>1\),
and in \(R\) we have \(q>190\).
There are two \(F\) branches for \(v\) and one \(R\) branch,
each defined throughout the small \(p\) interval.
In each,
\[
 \frac{dh}{du}=-\frac{p}{1-Y(u)^2}+L'(u)+L'(v)\frac{p}{q}.
\]
This is positive at \(p=-1/250\) and \(p=0\) and negative
at \(p=1/250\).
On this interval its derivative in \(u\) is strictly negative:
the coefficient of \(Y''(u)\) is
\(-1/(1-Y(u)^2)+L'(v)/q<-499/500\), and the remaining terms
have absolute sum less than \(1\) by~\eqref{eq:diagram-bounds}.
Thus each branch has one simple maximum, at \(0<p<1/250\).
Transposition yields the three simple minima in the disjoint
region where \(|q|\leq1/250\). This exhausts all critical points.

Call the two positive maxima on \(F\times F\) with
\(v>x_*\) and \(v<x_*\) the values \(a,b\) respectively,
and the maximum on \(F\times R\) the value \(c\).
Positivity follows from \(y=-Y(u)>1/250-p^2/400\)
and \(\abs{L(u)-L(v)}<4J_c<1/500\).
Holding \(u\) fixed and moving to a smaller solution \(v\)
of \(Y(v)=-Y(u)\) raises \(h\) since \(L'>0\) on the \(F\)
branch. Moving from the smaller \(F\) solution to the \(R\)
solution also raises \(h\), since \(L\) is positive on \(F\)
and negative on \(R\). This proves \(a<b<c\).
The rational interval enclosures in the proposition
are certified as described in Appendix~\ref{app:certificate}.

On \(R\times R\) the field increases strictly with \(y\)
and tends to \(\pm\infty\) at the corresponding ends.
On \(F\times R\) both ends tend to \(h=-\infty\)
and the unique turning point is the maximum \(c\).
The transposed component has the opposite behavior.
On the circle the four intersections with \(u=x_*\) or
\(v=x_*\) alternate in the sign of \(h\).
The circle has exactly two maxima and two minima, so
there are four roots at \(h=0\), one for each sign pair
\((\operatorname{sgn}p,\operatorname{sgn}q)\).
The two positive maxima successively remove the pair
with \(q>0\), then the pair with \(q<0\).
The fold root itself is counted once at a critical value.

To compute the indices, start at \(h=0\).
The eliminated degree-two directions and the endpoint
antisymmetric directions are positive. In the remaining
\((u,v,y)\) coordinates the Hessian is
\[
 \begin{pmatrix}
 A_1p&0&A_1\\
 0&A_2q&-A_2\\
 A_1&-A_2&d
 \end{pmatrix},
 \qquad A_i>0,\quad d\ge1 .
\]
Indeed the endpoint residual \(P(x,Y(x))=0\) has
\(P_y<0\) and \(P_x=-P_yY'(x)\). Summing the two endpoint
contributions per pair and reflecting the left pair
gives the first two rows; the control diagonal satisfies
\(d\ge1\) by~\eqref{eq:hessian}. At \(h=0\),
\(\abs y\le\tanh(4J_c)<1/500\), and the correlation bound
\(\beta\le\alpha\) for a control edge gives \(A_i<1/500\)
in the displayed coordinates. Here \(|p|,|q|>\sqrt{4/5}\)
for each \(F\) variable by strong convexity and
\(Y_*\!<-1/250\); for each \(R\) variable the derivative
is greater than \(190\).
The Schur complement \(d-A_1/p-A_2/q\) is positive.
The index is therefore the number of negative entries
among \(p,q\). Along a noncritical branch it is constant.
At a simple fold the two adjacent indices differ by one
and the fold has one zero eigenvalue and the smaller
negative index. This yields the two polynomials for each
row of Table~\ref{tab:one-field}.
\end{proof}

\subsection{A root-count formula for Euler characteristic}

For any fixed finite real couplings and fixed fields on \(W\),
use the positive coefficient coordinates \(b_D=e^{2h_D}\) for
the free fields. Let \(A_0\) be any semialgebraic subset of the
response and \(\pi_0\) its projection to \(b_D\).
By Theorem~\ref{thm:A} every fiber is finite.
Semialgebraic triviality and cell decomposition give a finite
partition of the field space into open cells \(\sigma\) over
which \(\pi_0\) is a product with a finite set of size \(n_\sigma\).
Additivity of compactly supported Euler characteristic gives
\begin{equation}
 \chi_c(A_0)=\sum_\sigma(-1)^{\dim\sigma}n_\sigma .
 \label{eq:root-euler}
\end{equation}
Every boundary cell in a chosen compatible decomposition
contributes according to its dimension. The counts \(n_\sigma\)
are distinct roots in \(A_0\). For a regular compact connected
surface component \(C\), one has
\(\chi_c(C)=\chi(C)=2-2g(C)\).
Away from critical values, its field projection is a proper
local analytic diffeomorphism and is a finite covering over
each connected region in its image. \cite[\S2.1]{hardt}

\subsection{The complete two-field product diagram}

Take the disjoint union of two copies of the nine-vertex model,
with fields \(h_1,h_2\) on the two control vertices.
The positive response is the product of the one-field responses.
Its only compact connected component is \(C=C_1\times C_1\).
The full discriminant is the union of the twelve lines where
one coordinate is in \(\{\pm a,\pm b,\pm c\}\).
The discriminant restricted to \(C\) consists of the segments
where one coordinate is in \(\{\pm a,\pm b\}\) and the other
is in \([-b,b]\).
At a field pair on two discriminant lines there is a root
with two simultaneous independent folds. The Hessian is a direct
sum of the Hessians from the two copies, so multiplying the
one-field polynomials gives all counts and inertias, including
on the boundary lines and at their intersections.

To recover the genus by~\eqref{eq:root-euler}, partition each
field axis by the six thresholds. The sum of the circle counts
on its point cells is \(2\cdot3+2\cdot1=8\).
The sum on its open intervals is \(4+2+2=8\).
The sums of the counts on \(C\) over two-, one-, and zero-dimensional
cells are then \(64,128,64\), respectively.
Thus \(\chi(C)=64-128+64=0\) and the genus is one.

The complete diagram is for this disconnected product graph.
A coupling between the two control vertices alone preserves the
conditional response and hence the torus, while changing its
field projection by the corresponding edge contribution.
The connected simple bipartite cubic construction in
Theorem~\ref{thm:C} gives regular surfaces of all genera
under its separate coupling and field constraints.

\section{Formal verification and reproducibility}
\label{sec:formal}

The three main theorems have been formally verified in Lean~4 with
Mathlib. This section records what was verified, where the formal
proofs depart from the written ones, how the development is audited and
rebuilt, and which parts of the paper remain outside it. The Lean
sources are distributed in the accompanying repository
\cite{responselean}; all statements below refer to its revision
\texttt{\leancommit}.

\subsection{Verified statements}

Table~\ref{tab:lean} lists the final declarations. All modules lie in
the namespace \texttt{BetheGeometry.Response}.

\begin{table}[htbp]
\centering
\footnotesize
\begin{tabular}{@{}>{\raggedright\arraybackslash}p{3.4cm}>{\raggedright\arraybackslash}p{7.0cm}>{\raggedright\arraybackslash}p{5.0cm}@{}}
\toprule
Result & Final declaration & Module \\
\midrule
Theorem~\ref{thm:A} & \texttt{finite\_admissibleSolutions\_ising} & \texttt{UniversalFiniteness} \\
Theorem~\ref{thm:B}, (i)--(iii) & \texttt{theoremB\_smooth} & \texttt{TheoremB} \\
Theorem~\ref{thm:B}, minors & \texttt{theoremB\_smooth\_simple} & \texttt{TheoremB}, \texttt{ForbiddenMinors} \\
Theorem~\ref{thm:C}, realization & \texttt{theoremC\_genus} & \texttt{TheoremC} \\
Theorem~\ref{thm:C}, size bounds & \texttt{theoremC\_minVertices} & \texttt{TheoremC} \\
Morse inequality used for the lower bound & \texttt{polynomialSurfaceMorseInequality} & \texttt{CircuitChainSize}, \texttt{PolynomialMorseBound} \\
Genus witness has \(b_0=b_2=1\), \(b_1=2g\) & \texttt{SurfaceGenusWitness.rationalBetti\_zero}, \newline \texttt{rationalBetti\_one}, \texttt{rationalBetti\_two}, \texttt{surfaceGenus\_eq} & \texttt{SurfaceGenusHomology} \\
\bottomrule
\end{tabular}
\caption{Final Lean declarations for the main theorems at revision
\texttt{\leancommit}. The modules are files under
\texttt{BetheGeometry/Response/} in the repository.}
\label{tab:lean}
\end{table}

\emph{Theorem~\ref{thm:A}.} The final statement takes a finite loopless
multigraph with edge identities, arbitrary real couplings
\(J\colon E\to\R\) and fields \(h\colon V\to\R\), and proves that the
type of admissible complex fixed points, defined by~\eqref{eq:bp}
and~\eqref{eq:admissible} with parallel edges as distinct factors, is
finite. Zero couplings are removed through an explicit bijection with
the fixed points of the reduced graph.

\emph{Theorem~\ref{thm:B}.} The declaration \texttt{theoremB\_smooth}
proves the equivalence of the three conditions as stated, for loopless
multigraphs. The global real analytic coordinate of condition (ii)
includes disconnected graphs, and its inverse extends locally to the
ambient space. The declaration \texttt{theoremB\_smooth\_simple}
connects the classification to the three forbidden minors for simple
graphs, with minors defined through branch sets.

\emph{Theorem~\ref{thm:C}.} The declaration \texttt{theoremC\_genus}
states that for every \(k\geq1\) the wired graph is connected, simple,
bipartite, and cubic with \(366k+2550\) vertices, two distinct controls,
and coupling \(J_0\) on every edge, and that for every
\(0\leq g\leq3\cdot2^k-1\) there are noncontrolled fields bounded by one
constant \(H_0\), fixed before \(k\) and \(g\), such that the response to
the two remaining fields has a regular compact connected component
carrying a genus-\(g\) witness. Genus is formalized as a connected-sum
presentation: the witness exhibits the component as homeomorphic to an
iterated connected sum of standard spheres and tori along collared chart
disks. The module \texttt{SurfaceGenusHomology} proves that every
genus-\(g\) witness has rational Betti numbers \(b_0=b_2=1\) and
\(b_1=2g\), with no additional manifold or orientation hypothesis,
identifies the witness integer with the homological genus
\(\lfloor b_1/2\rfloor\) of the Tau Ceti library \cite{tauceti}, and
proves the uniqueness of that integer and its invariance under
homeomorphism. The declaration \texttt{theoremC\_minVertices} gives one
field bound \(H_0>0\) under which every genus is realizable and
\[
 \frac{\log(g+1)}{6\log5}-\frac13
 \ \leq\ n_{\min}(g)\ \leq\ 366\,k(g)+2550,
 \qquad
 k(g)=\max\bigl(1,\lceil\log_2((g+1)/3)\rceil\bigr),
\]
with no hypothesis beyond the standard axioms. The minimality of
\(k(g)\) and the bound \(k(g)\leq\log_2(g+1)+1\) are verified as well.

\subsection{Where the formal proofs differ from the written proofs}

\emph{Theorem~\ref{thm:A}.} The formal proof follows
Section~\ref{sec:finiteness} but replaces the projective solution curve
by a finitely generated function field with a valuation that need not be
discrete. The stationarity of the rational invariant, the regularizing
spin flip, the pole budget, and the edge excess of every pole layer are
proved in that setting.

\emph{Theorem~\ref{thm:B}.} The forward direction excludes circular
components without the planar embedding argument of
Section~\ref{sec:rigidity}, by following the sign of the sweep
derivative at the lowest point of each height. On the converse side, the
formal proof obtains the two turning arcs from three zero-field roots on
each connected core of cycle rank two, by the Bethe-potential route
described in Section~\ref{sec:rigidity}, and constructs a compact
connected component on which the sweep derivative is surjective after
perturbing the couplings. The canonical tangent traverses this component
periodically, so the time coordinate modulo its least period gives a
smooth circular parametrization with a smooth local inverse.

\emph{Theorem~\ref{thm:C}.} The formal route differs from
Sections~\ref{sec:topology} and~\ref{sec:uniform} in two respects.
First, all cuts are made on the exact iterated double, the system
\(Z_j^2+Z_{j-1}=c_j\) over the circuit torus, where the removed disks are
explicit chart disks joined by explicit corridor charts, so no
Morse-theoretic identification of excursion sets is needed. A handle
attached by a two-disk surgery along a corridor chart is identified with
a torus summand. The cubic chain is rescaled by \(U_j=t^jZ_j\), so that
the mixing coefficient between the folds \(j\) and \(j+1\) is
\(t^{j+2}\) (indices from \(0\)); the rescaled system converges to the
exact iterated double as \(t\to0\), and one persistence step followed by
the transfer to the wired BP graph gives the component. Second, the size
lower bound is obtained from the algebraic bound \(5^{6n+2}\) on
nondegenerate critical points of a linear height on the response of a
cubic graph with two controls, together with the existence of a Morse
linear height, rather than from the sharper polar-degree estimate of
Section~\ref{sec:uniform}. This coarser exponential estimate suffices
for the logarithmic bound. The Morse inequality used,
\texttt{polynomialSurfaceMorseInequality}, states that a Morse linear
height on a regular compact component of a polynomial constraint system
in the positive orthant, carrying a genus-\(g\) witness, has at least
\(2g\) critical points; this weaker form suffices. Its proof passes
through rational singular homology: a Mayer--Vietoris argument along the
chart disks shows that the Betti numbers of the component are finite,
vanish above degree two, and satisfy \(b_0-b_1+b_2=2-2g\), hence
\(b_0+b_1+b_2\geq2g\); the component is identified with a regular level
set of the constraint map, on which the height is a Morse function in
the sense of Tau Ceti, whose Morse inequalities bound the sum of the
Betti numbers by the number of critical points.

\subsection{Reused libraries and new development}

From Mathlib the development uses singular homology with rational
coefficients, the implicit and inverse function theorems, analytic
functions, polynomial rings, valuations, and the spectral theorem for
symmetric real matrices. From the Tau Ceti library \cite{tauceti} it
uses the Morse inequalities for closed manifolds, the Mayer--Vietoris
sequence for rational Betti numbers, the homology of spheres and tori,
and the charted-space structure of regular level sets. From the
formalization \texttt{BPFixedPoints} of \cite{counting}, cited there as
the accompanying repository, it reuses the BP equations, the admissible
locus, the Eulerian sign sectors, the removal of zero couplings, and the
uniqueness and count of fixed points at strong fields. Everything specific to this paper is new: the
valuation argument of Theorem~\ref{thm:A}, the analytic response line
and the exclusion of periodic orbits, the two-fold circle and the
coupling perturbation, the forbidden minors, the \(K_{2,3}\) fold and
its retuning, the transmission and cancellation blocks and their
stationary elimination, the wiring of the chain, the scaling limit to
the cubic circuit, the iterated double with its corridor charts and
connected-sum witness, the rational homology of the witness, the
polynomial critical-point bound, and the final theorems.

\subsection{Axiom audit and build}

An axiom audit, \texttt{ResponseAudit}, runs as part of the default
build. It inspects every declaration whose name lies in the namespace
\texttt{BetheGeometry.Response}, including private and auxiliary
declarations, together with its transitive dependencies, and fails
unless the only axioms used are \texttt{propext},
\texttt{Classical.choice}, and \texttt{Quot.sound}. In particular no
declaration depends on \texttt{sorryAx} or on \texttt{native\_decide}.
At revision \texttt{\leancommit} the audit reports \(6709\)
declarations and \(5452\) theorems. The audit certifies this
development, not every file of the repository.

The toolchain is \texttt{leanprover/lean4:v4.36.0-rc2}. Mathlib is
pinned to commit \texttt{1b1ccf4b}, Tau Ceti to revision
\texttt{e583e15d}. The Lean 4.36-compatible modules from
\texttt{BPFixedPoints} are included under \texttt{vendor/BPFixedPoints},
with their source revision documented there. The build is
\begin{verbatim}
lake exe cache get
lake build
\end{verbatim}
from the root of the Lean project. The first command obtains the pinned
Mathlib cache and requires network access; the second checks the whole
library and runs the audit.

\subsection{What remains outside the formal development}

The following parts of the paper are not part of the verified
statements. Corollary~\ref{cor:budget} and the sector parity of the
length sums. Proposition~\ref{prop:zero-field} beyond the three
zero-field roots used in the converse of Theorem~\ref{thm:B}.
Proposition~\ref{prop:nash} and the cactus example with \(2N+1\)
crossings. Lemma~\ref{lem:persistence} is verified in the form used by
the formal proofs, with the persisted component homeomorphic to the
original one, rather than with the diffeomorphism and \(C^q\) closeness
stated there. Lemma~\ref{lem:negative-index} beyond its linear algebra, that
is, the location of the negative modes on the constructed surfaces. The
polar-degree estimate of Section~\ref{sec:uniform}, which is replaced by
the coarser bound described above. Section~\ref{sec:diagrams}, whose
certificates are checked by the separate interval arithmetic code
described in Appendix~\ref{app:certificate}; the formal development
does not contain that example.

\subsection{Certified example}

The rational interval certificates of Section~\ref{sec:diagrams} are
produced and checked by the experiments \texttt{EXP-215} and
\texttt{EXP-324} distributed with the accompanying repository
\cite{responselean}. Each experiment records its inputs, code, saved
results, and the hashes of its sources, and a single command re-runs
the check against the saved results (Appendix~\ref{app:certificate}).
These certificates are exact rational computations and are independent
of the Lean development.

\subsection{Scope of the results}

The theorems are statements about the geometry of fixed points, and the
following limits apply. Theorem~\ref{thm:A} isolates every admissible
root; it does not supply an algorithm or a bound on the cost of finding
roots. Theorem~\ref{thm:B} classifies the global shape of the full
positive response under a positive-direction sweep; several positive
roots at one field are permitted on a single line. Theorem~\ref{thm:C}
realizes every genus on surfaces of Bethe saddles, and \(n_{\min}(g)\)
measures vertices only, not the precision of the fields or the
conditioning of the response.

\section*{Declarations}
Large language model tools were used in exploring candidate statements,
in drafting, reorganizing, and checking the proofs and the text, in
writing the Lean formalization, and in preparing the interval
certificates of Section~\ref{sec:diagrams}. The author assumes
responsibility for all content, including the definitions, the written
proofs, the correspondence between the written and the formal
statements, and the citations.

\appendix
\section{The Nash one-state representation}
\label{app:nash}

Fix \(p_U(G)\leq1\) and write \(r_n=1\).
With \(v=b_n/(1+b_n)\) and \(P_\alpha\) the product of the
incoming message ratios at the tail excluding the reverse, use
the equations
\[
 y_\alpha\bigl(r_i v P_\alpha+a_e(1-v)\bigr)
   -\bigl(a_e r_i v P_\alpha+1-v\bigr)=0,\qquad 0\leq v\leq1 .
\]
For positive messages these imply
\(a_e^{-1}\leq y_\alpha\leq a_e\). At \(v=0\) the only
solution is \(y_\alpha=a_e^{-1}\), and at \(v=1\) it is
\(y_\alpha=a_e\). The message Jacobians at these two ends are
diagonal, with entries \(a_e\) and \(r_i P_\alpha\), respectively.
By the implicit function theorem and the uniform message box
on compact subsets of the base, these are the two boundary
sheets of a Nash manifold with boundary. Projection to \((a,r)\)
is proper by the same box and \(0\leq v\leq1\).

In the interior, \(D_m\log r=2RK\), where the rows of \(R\) are
\(e_i^\top-e_n^\top\). It has full row rank because the kernel
of a singular \(K\) is spanned by a nonnegative vector with
positive coordinate sum. On a disconnected graph only the
one component outside the pseudoforest class can have a kernel.
At the boundary the endpoint charts show submersivity, with
each boundary sheet mapping isomorphically onto the base.

The base is Nash-diffeomorphic to Euclidean space, using
\(x\mapsto x-1/(x-1)\) on \((1,\infty)\) and
\(x\mapsto x-1/x\) on \((0,\infty)\).
At a fixed base point with all couplings sufficiently weak the
fiber is Nash-isomorphic to \([0,1]\) via \(v\): contraction gives
unique simple roots for all interior \(v\), and the boundary
charts extend the parametrization. The boundary version of
proper Nash submersion triviality gives a product with \([0,1]\)
over the whole base, with the two sheets corresponding to
the two ends. The zero-dimensional base of a single vertex is
immediate from \(v\) itself. Restricting to the interior gives
the state \(s\in(0,1)\) and the claimed scalar function
\(\mathcal B=v/(1-v)\). Here we apply the theorem with no marked
interior submanifolds and with the boundary restriction also a
proper onto submersion, as checked above. Translate the base
coordinate so that the weak-coupling base point is its origin.
The theorem fixes that fiber pointwise, and hence identifies
its state with \(v\). Connectedness of each boundary sheet then
keeps the endpoints \(v=0,1\) in that order over the whole base.
\cite[Theorem~I, pp.~1401--1402]{nash}

Along a state fiber the oriented tangent in magnetizations is a
positive multiple \(c\adj(K)\ones\), by the end orientation and
\eqref{eq:tangent}. Differentiation gives
\(\partial_s\mathcal B=2\mathcal B c\det K\).
The critical points are finite for each base point. Their number
is uniformly bounded over the base for fixed \(G\) by semialgebraic
triviality applied to the critical set. When \(p_U\geq2\),
Theorem~\ref{thm:B} supplies a circular component in a uniform
sweep, which precludes a product preserving \((a,r)\).

\section{Fold blocks, wiring, and weak-coupling scaling}
\label{app:wiring}

\subsection{Numerical separation in the seed}

The two stationary \(b\) values for \(f_L\) are
\(b_\pm=(L\pm\sqrt{L^2+8})/4\).
For \(L=1/8,1/16\) the two positive heights at local maxima are
\[
 h^-(L)=\sqrt{1-b_+^2}\,(b_+-L),\qquad
 h^+(L)=\sqrt{1-b_-^2}\,(L-b_-).
\]
For \(L=1/8\) they are in \((0.413,0.415)\) and
\((0.589,0.591)\), respectively. For \(L=1/16\) they are in
\((0.456,0.457)\) and \((0.544,0.546)\), respectively.
These rational brackets follow by bounding the positive square
roots between rationals by squaring. The four sums give the
separations in Section~\ref{sec:topology}. Each mixed critical
value is the difference of two of the bracketed heights, with
absolute value less than \(1/4\).

\subsection{The local folds and their port coordinates}

Let \(J_*=\tfrac14\log 10\), \(t_*=\tanh J_*\). On a pair of
vertices with three separate factors of coupling \(J_*\), the
equal-magnetization field \(f(x)\) has derivative
\begin{equation}
 f'(x)=\frac{3/\sqrt{10-9x^2}-1}{2(1-x^2)}.
 \label{eq:f-derivative}
\end{equation}
Indeed, for equal endpoint magnetizations \(x\) the edge odds
ratio gives correlation \((S-1)/(S+1)\), \(S=\sqrt{10-9x^2}\).
The common-direction row sum of~\eqref{eq:hessian} is then
\((3/(1+\beta)-2)/(1-x^2)\). The two folds occur at \(x=\pm1/3\),
with \(f''(\pm1/3)=\pm3/16\) and
\(f(1/3)=\tfrac14\log(125/128)=-f(-1/3)\).
For the values of \(f\) itself, write
\(\xi=(11-2\sqrt{10-9x^2})/9\) for the edge spin product mean.
The ratio of the \(++\) to the \(--\) entry of the edge table
is the fourth power of the exponential of either cavity field.
Thus
\begin{equation}
 f(x)=\frac34\log\frac{1+2x+\xi}{1-2x+\xi}-2\atanh x.
 \label{eq:core-field}
\end{equation}
At \(x=1/3\) the ratio in the logarithm is \(5/2\), giving
the stated field.

Replace a parallel factor of coupling \(J_*\) by a path of two edges
with \(\tanh J_{\rm path}=\sqrt{t_*}\), giving its internal vertex
zero field. Eliminating the internal messages gives the same
endpoint update, since \(g_s\circ g_t=g_{st}\). The added
stationary directions are positive by the principal-block property
of~\eqref{eq:hessian}. Replacing two of the factors gives a diamond,
and replacing all three gives \(K_{2,3}\) with its degree-two
vertices available for attachments. The antisymmetric endpoint
direction is positive by~\eqref{eq:hessian} on the two-vertex
reduction, so each block has exactly one kernel direction at its fold.

Use a state coordinate \(s\) increasing with \(x\) and
\(f(x)=f(x_*)+\sigma s^2\), where \(x_*\) is the selected fold.
The endpoint derivative at \(s=0\) is \(d_*=\sqrt{32/3}\).
On a two-edge path the middle spin has conditional expectation
\(\tanh(2J_{\rm path})(\sigma_i+\sigma_j)/2\) at zero internal field.
Thus along the symmetric branch each port has magnetization
\(\gamma x\), with \(\gamma=\tanh(2J_{\rm path})=3/\sqrt{10}\)
and derivative \(\gamma d_*\) in \(s\). The cubic term of the
block potential after eliminating the complementary variables is
\(2d_*\sigma s^3/3\). Indeed, along the symmetric curve the
gradient is zero on the internal vertices and is
\(f(x)-f(x_*)=\sigma s^2\) at both endpoints. Differentiating
along the curve multiplies by \(2(d_*+O(s))\). An equal
endpoint field shift \(\delta h\) contributes the linear term
\(-2d_*\delta h\,s\).

\subsection{Wiring and the scaling limit}

Number the four circle blocks \(0,1,2,3\) and the \(k\) added
folds \(4,\ldots,p-1\), where \(p=k+4\).
The edges of the cubic potential are the two links \(0\)--\(1\)
and \(2\)--\(3\), the six links from the controls to
\(\{0,1,4\}\) and \(\{2,3,4\}\), respectively, and the \(k-1\)
links of the chain starting at block \(4\). Thus there are \(p+3\)
logical edges. For a \(K_{2,3}\) block use a separate port for
each incident logical edge. The two control vertices each have
degree three. Exactly \(p\) ports are unused because the number
used is \(2(k+1)+6\). The weak-coupling graph has \(5p+2\)
vertices and \(7p+3\) edges. In the diamond implementation each
block link is a pair of matching edges between its endpoints
and each control link uses one edge to each endpoint.
The seed has \(22\) vertices and \(41\) edges, and each added
fold increases these counts by \(4\) and \(7\).

For the \(K_{2,3}\) implementation, put coupling
\(2\lambda_{ij}\rho^3/(\gamma^2d_*)\) on a port link and
\(2a_{i\alpha}\rho^4/\gamma\) on a port-to-control link,
with \(\rho>0\) small. In the diamond implementation use
\(\lambda_{ij}\rho^3/d_*\) on each matching edge and
\(a_{i\alpha}\rho^4\) on each control edge. Absent logical edges
are omitted. Let \(m^*\) consist of the chosen fold magnetizations
with control magnetizations zero. Use the fields
\(H_{J,W}(m^*)+\rho^6\sum_i t_i e^{(i)}\), where \(e^{(i)}\)
has entries one at the two endpoints of block \(i\) and zero
elsewhere. The field map includes the newly attached edges.

For one weak edge, \(L_e-S_i-S_j\) is analytic near \(J=0\),
and its derivative in \(J\) at zero is \(-m_i m_j\).
After subtracting its reference value and linear terms in the
magnetization displacement, the edge contributes
\[
 -J\,\delta m_i\delta m_j+
 O\bigl(J^2(\abs{\delta m_i}+\abs{\delta m_j})^2\bigr).
\]
Scale \(s_i=\rho^3 X_i\) and \(m_{c_\alpha}=\rho^2Y_\alpha\).
The complementary magnetizations are solved by the implicit
function theorem and are \(O(\rho^6)\) relative to the symmetric
block curve. This follows by scaling their displacement by
\(\rho^6\) and dividing their equations by \(\rho^6\), so that
the limiting derivative is the positive complementary Hessian.
The displacement and its state derivatives have these bounds
on fixed compact coordinate sets.

Subtract the terms depending only on \(Y\) and constants from
the reduced potential, and divide by \(2d_*\rho^9\).
The block cubic, the weak block edges, the control edges,
and the field offsets now give the four terms of~\eqref{eq:circuit},
respectively. The energy correction from the complementary
variables, fourth and higher block terms, and quadratic and
higher corrections in the weak couplings have weight at least
\(\rho^{12}\) before division. Taylor expansion on fixed compact
sets gives \(C^q\) convergence to~\eqref{eq:circuit} with error
\(O(\rho^3)\) for each fixed finite \(q\).
All reference fields are finite and the fold endpoint
fields above have magnitude below one. Hence the couplings
and noncontrolled fields of these weak-edge models can be
taken to have magnitude less than one.

\section{Uniform-coupling block calculations}
\label{app:uniform}

\subsection{Stable transmission}

Use interior vertices \(A,B\) in opposite bipartition classes
and two disjoint length-three paths \(A-p_\pm-q_\pm-B\).
For each sign, take a copy of \(K_{3,3}\) with one edge
deleted and connect its two deficient vertices to \(p_\pm,q_\pm\)
of the opposite colors, respectively. Each of \(A,B\) is joined
by one edge to an exterior terminal. The interior vertex count
is \(2+4+12=18\), and all interior vertices are now cubic.
Take zero magnetization at \(A,B\) and at the exterior
terminals. Assign magnetization \(s\) to all eight vertices in
the positive half and \(-s\) to all eight in the negative half,
with \(0<s<1\).

For a binary edge, its squared correlation is the product
of the two regression slopes of one spin on the other.
A slope has the form \(t(1-w^2)/(1-t^2w^2)\), where \(w\)
is the relevant cavity magnetization. It is at most \(t\).
If one full magnetization tends to saturation, the
corresponding cavity field differs from the full field by
at most \(J_0\), so its slope tends to zero. Therefore as
\(s\to1\) every edge correlation in the block tends to
zero except those of the two exterior links, which are
\(t\). The normalized interior Hessian tends to a positive
diagonal matrix, with entries \(1/(1-t^2)\) at \(A,B\).
At an exterior terminal the remaining limiting diagonal
contribution cancels under elimination:
\[
 \frac{t^2}{1-t^2}
 -\frac{(t/(1-t^2))^2}{1/(1-t^2)}=0 .
\]
The limiting mixed entry is zero. For finite \(s\) sufficiently
close to \(1\) the interior block is positive definite with
strictly positive inverse entries, since it is an irreducible
symmetric matrix with nonpositive off-diagonals. The two
negative exterior-to-interior coefficients give a strictly
negative mixed entry after elimination. Exchanging the two
halves and reversing spins preserves the internal fields
and yields evenness of the exterior contribution.

\subsection{Crossing and cubic coefficient of the three-port block}

In \(K_{2,3}\) take the three ports at zero magnetization.
For either same-sign or opposite-sign magnetizations of the
internal pair, each edge has the same positive correlation
\(b\). Let \(w\in[0,1)\) be the absolute cavity magnetization
at an internal vertex on one such edge. The conditional means
given the two port spins yield
\[
 x=\frac{w(1-t^2)}{1-t^2w^2},\qquad
 b^2=\frac{t^2(1-w^2)}{1-t^4w^2}.
\]
The first expression increases and the second decreases
strictly for \(w\in(0,1)\).
Also \(t^2=(\sqrt{10}-1)/(\sqrt{10}+1)\in(1/2,5/9)\).
The internal-pair difference direction and the two
sum-zero port directions are positive Hessian directions.
On the normalized subspace constant on each vertex class,
multiplying the Hessian by \(1-b^2\) gives
\[
 \begin{pmatrix}
    1+2b^2 & -\sqrt6\,b\\
    -\sqrt6\,b & 1+b^2
 \end{pmatrix}.
\]
Its determinant is \((1-b^2)(1-2b^2)\). The lowest eigenvalue
crosses simply from negative to positive at \(b^2=1/2\).
The remaining eigenvalues are positive. Small diagonal
corrections at the ports preserve this simple crossing.

At \((x_f,x_f)\) and the crossing, write \(v=1-x_f^2\).
On one edge, use centered standardized spins \(X\) at
the internal endpoint and \(Y\) at the port. Then
\[
 \mathbb E X^3=M=-2x_f/\sqrt v,\quad
 \mathbb E X^2Y=Mb,\quad \mathbb E XY^2=\mathbb E Y^3=0 .
\]
Use fixed linear normalized coordinates at this reference point,
so a normalized tangent \(q\) represents the physical magnetization
direction \(\diag(\sqrt{1-m_i^2})q\). Choose the normalized
kernel vector to have coordinate \(1\)
at each internal vertex and \(4b/3\) at the ports. On an
edge the inverse normalized covariance applied to this
vector gives coefficients \((2/3,2b/3)\).
The third derivative of the Legendre potential of the edge
in this direction is minus the third cumulant of
\((2/3)X+(2b/3)Y\). This follows by differentiating the
inverse covariance formula for its Hessian. The derivative
is \(-20M/27\). For each internal vertex the single-vertex
negative entropy has derivative \(-M\) in the same normalized
direction; the port entropy has derivative zero. Summing
over six edges and subtracting the \((\deg-1)\) vertex
terms gives
\[
 6(-20M/27)+4M=-4M/9>0.
\]
Spin reversal gives the negative value at the opposite fold.
The nonzero signs persist for small port corrections.
The ancillary terms used in Section~\ref{sec:uniform}
are even at zero port magnetization and contribute no
third derivative there.

\section{Rational interval certificate for the product example}
\label{app:certificate}

The interval inputs and code are the experiments
\texttt{EXP-215} and \texttt{EXP-324} of the accompanying repository
\cite{responselean}. The certificates use
outward-rounded rational intervals on a \(2^{-96}\) grid.
The square root of a positive interval is enclosed by taking
an integer square-root lower bound and rounding the upper
bound upward, with containment checked by squaring. All formulas
below can be evaluated without logarithms.

Set \(S=\sqrt{10-9x^2}\), \(\xi=(11-2S)/9\) and
\(P_\pm=1\pm2x+\xi\). In the notation of
Section~\ref{sec:diagrams},
\[
 E(x)=e^{2f(x)}
    =\left(\frac{P_+}{P_-}\right)^{3/2}
       \left(\frac{1-x}{1+x}\right)^2,\qquad
 z_0(x)=\frac{E(x)-e^{2\eta}}{E(x)+e^{2\eta}},
 \qquad e^{2\eta}=\sqrt{100001/102400}.
\]
The five endpoint and central tests are
\[
 \begin{gathered}
 \tau(-61/100)<-1,\qquad \tau(-59/100)>1,\\
 \tau(1/4)>1,\qquad \tau(41/100)>1,\qquad -1<\tau(1/3)<0 .
 \end{gathered}
\]
Since \(f\) increases for \(|x|>1/3\) and decreases for
\(|x|<1/3\), with limits \(\pm\infty\) at \(\pm1\),
these isolate the two allowed intervals used in the text.
At \(\tau=\pm1\), the limiting value of \(Y\) is \(\pm1\).

The following enclosures are checked on each of the \(128\)
equal closed subintervals of the indicated boxes, using
\eqref{eq:f-derivative} and its derivative.
\[
 \begin{array}{c|cc}
 \text{Box}& f'&f''\\ \hline
 \left[-61/100,-59/100\right]&(1/10,\,3/20)&(-2,\,2)\\
 \left[1/4,41/100\right]&(-1/40,\,1/40)&(11/100,\,1/2)
 \end{array}
\]
The derivative of~\eqref{eq:f-derivative} is
\(27x/[2 S^3(1-x^2)]+(3/S-1)x/(1-x^2)^2\).
For \(Y=G(\tau,x)\), the identities for the chain rule are
\[
 \tau'=(1-z_0^2)f'/\alpha,\qquad
 \tau''=(1-z_0^2)(f''-2z_0(f')^2)/\alpha,\qquad
 Y''=G_\tau\tau''+G_{\tau\tau}(\tau')^2+2G_{\tau x}\tau'.
\]
Here \(G\) is affine in \(x\), and for \(|\tau|\le1\),
\[
 G_\tau=\frac{1-\alpha^2}{2}
 \left(\frac{1+x}{(1+\alpha\tau)^2}
       +\frac{1-x}{(1-\alpha\tau)^2}\right).
\]
It lies between \(99/100\) and \(101/100\), with
\(0\le G_x\le\alpha\) and
\(\max(|G_{\tau\tau}|,|G_{\tau x}|)<1/500\).
Substitution gives the bounds on \(Y\) derivatives in
\eqref{eq:diagram-bounds}. For the bounds on \(L\), write
\(A=\atanh x-f(x)+\eta\) and \(w=\tanh A\).
Then \(L=2\atanh(\alpha w)\), \(0<A'<17/10\),
and \(|A''|<6\) on both boxes. It follows that
\(w'>0\), \(|w''|<12\), and
\(0<L'<1/500,\ |L''|<1/50\).
Since \(L(0)\) need not be evaluated, the sign of \(L\)
on the allowed intervals follows directly from the
numerator \(x-z_0(x)\), which has the sign of \(x\)
there.

On \(I=[8/25,26/75]\), \texttt{EXP-215} uses \(256\)
equal closed intervals to check \(|\tau|<1\) and
\(Y''\geq200\). It also checks \(Y>1/50\) at both
endpoints of \(I\), and
\[
 -1/200<Y(1/3)<-1/250,\qquad |Y'(1/3)|<1/1000 .
\]
The minimum on \(I\) is interior and is greater than
\(Y(1/3)-Y'(1/3)^2/400>-1/190\), by strong convexity.
The derivative bounds on \(F\) make it the minimum there.

To enclose the three positive critical values,
use \(0<Y'(u)<1/250\), the center derivative bound, and
\(Y''>200\) to get \(|u-1/3|<1/40000\).
Taylor's inequality with \(Y''<1100\) encloses \(-Y(u)\)
using the center value. Evaluate \(Y\) at the ends of the
following intervals for \(v\):
\[
 [34/100,35/100]\ (a),\qquad [32/100,33/100]\ (b),\qquad
 [-601/1000,-599/1000]\ (c).
\]
In each case the interval lies on the specified monotone
branch, and the endpoint values bracket \(-Y(u)\).
For the field enclosure use
\[
 \tanh L(x)=\frac{2k(x)}{1+k(x)^2},
 \qquad k(x)=\alpha\frac{x-z_0(x)}{1-xz_0(x)} .
\]
The addition formula for \(\tanh\) then encloses
\(\tanh h\) in~\eqref{eq:diagram-curve}, giving the three
disjoint intervals in Proposition~\ref{prop:one-field}.
The arithmetic certifies derivative and threshold bounds.
The exact classification of components, root counts and
indices follows from those bounds and the proof in the text.

From the root of that repository, the saved results and source
hashes can be checked with
\begin{verbatim}
make verify EXP=EXP-215
make verify EXP=EXP-324
\end{verbatim}

\end{document}